\documentclass[aps,prx,twocolumn,longbibliography,superscriptaddress,noeprint,reprint]{revtex4-1}

\usepackage{graphicx}
\usepackage{mathtools}
\usepackage{tikz}
\usepackage{siunitx}
\usepackage{xcolor}
\usepackage[normalem]{ulem}
\usepackage{hyperref}

\usepackage[english]{babel}

\usepackage{float}
\usepackage{multirow}
\usepackage{sidecap}

\usepackage{xcolor}
\hypersetup{
    colorlinks,
    linkcolor={red!50!black},
    citecolor={blue!50!black},
    urlcolor={blue!80!black}
}

\usepackage{color}
\definecolor{blue}{rgb}{0.00,0.00,0.95}
\AtBeginDocument{%
    \newwrite\bibnotes
    \def\bibnotesext{Notes.bib}
    \immediate\openout\bibnotes=\jobname\bibnotesext
    \immediate\write\bibnotes{@CONTROL{REVTEX41Control}}
    \immediate\write\bibnotes{@CONTROL{%
    apsrev41Control,author="08",editor="1",pages="1",title="0",year="1"}}
     \if@filesw
     \immediate\write\@auxout{\string\citation{apsrev41Control}}%
    \fi
}%

\renewcommand{\d}{{\rm d}}

\begin{document}

\title{Geometry adaptation of protrusion and polarity dynamics in confined cell migration}

\author{David B. Br\"uckner}
\affiliation{Arnold Sommerfeld Center for Theoretical Physics and Center for NanoScience, Department of Physics, Ludwig-Maximilian-University Munich, Theresienstr. 37, D-80333 Munich, Germany}
\affiliation{Institute of Science and Technology Austria, Am Campus 1, 3400 Klosterneuburg, Austria}
\author{Matthew Schmitt}
\affiliation{Arnold Sommerfeld Center for Theoretical Physics and Center for NanoScience, Department of Physics, Ludwig-Maximilian-University Munich, Theresienstr. 37, D-80333 Munich, Germany}
\author{Alexandra Fink}
\affiliation{Faculty of Physics and Center for NanoScience, Ludwig-Maximilian-University, Geschwister-Scholl-Platz 1, D-80539 Munich, Germany}
\author{Georg Ladurner}
\affiliation{Faculty of Physics and Center for NanoScience, Ludwig-Maximilian-University, Geschwister-Scholl-Platz 1, D-80539 Munich, Germany}
\author{Johannes Flommersfeld}
\affiliation{Arnold Sommerfeld Center for Theoretical Physics and Center for NanoScience, Department of Physics, Ludwig-Maximilian-University Munich, Theresienstr. 37, D-80333 Munich, Germany}
\author{Nicolas Arlt}
\affiliation{Arnold Sommerfeld Center for Theoretical Physics and Center for NanoScience, Department of Physics, Ludwig-Maximilian-University Munich, Theresienstr. 37, D-80333 Munich, Germany}
\author{Edouard Hannezo}
\affiliation{Institute of Science and Technology Austria, Am Campus 1, 3400 Klosterneuburg, Austria}
\author{Joachim O. R\"adler}
\affiliation{Faculty of Physics and Center for NanoScience, Ludwig-Maximilian-University, Geschwister-Scholl-Platz 1, D-80539 Munich, Germany}
\author{Chase P. Broedersz}
\affiliation{Arnold Sommerfeld Center for Theoretical Physics and Center for NanoScience, Department of Physics, Ludwig-Maximilian-University Munich, Theresienstr. 37, D-80333 Munich, Germany}
\affiliation{Department of Physics and Astronomy, Vrije Universiteit Amsterdam, 1081 HV Amsterdam, The Netherlands}

\begin{abstract}
Cell migration in confining physiological environments relies on the concerted dynamics of several cellular components, including protrusions, adhesions with the environment, and the cell nucleus. However, it remains poorly understood how the dynamic interplay of these components and the cell polarity determine the emergent migration behavior at the cellular scale. Here, we combine data-driven inference with a mechanistic bottom-up approach to develop a model for protrusion and polarity dynamics in confined cell migration, revealing how the cellular dynamics adapt to confining geometries. Specifically, we use experimental data of joint protrusion-nucleus migration trajectories of cells on confining micropatterns to systematically determine a mechanistic model linking the stochastic dynamics of cell polarity, protrusions, and nucleus. This model indicates that the cellular dynamics adapt to confining constrictions through a switch in the polarity dynamics from a negative to a positive, self-reinforcing feedback loop. Our model further reveals how this feedback loop leads to stereotypical cycles of protrusion-nucleus dynamics that drive the migration of the cell through constrictions. These cycles are disrupted upon perturbation of cytoskeletal components, indicating that the positive feedback is controlled by cellular migration mechanisms. Our data-driven theoretical approach therefore identifies polarity feedback adaptation as a key mechanism in confined cell migration.
\end{abstract}

\maketitle

%%%%%%%%%%%%%%%%%%%%%%%%%%%%%%%%%%%%%
%%FIGURE 
\begin{figure*}[ht]
	\includegraphics[width=0.9\textwidth]{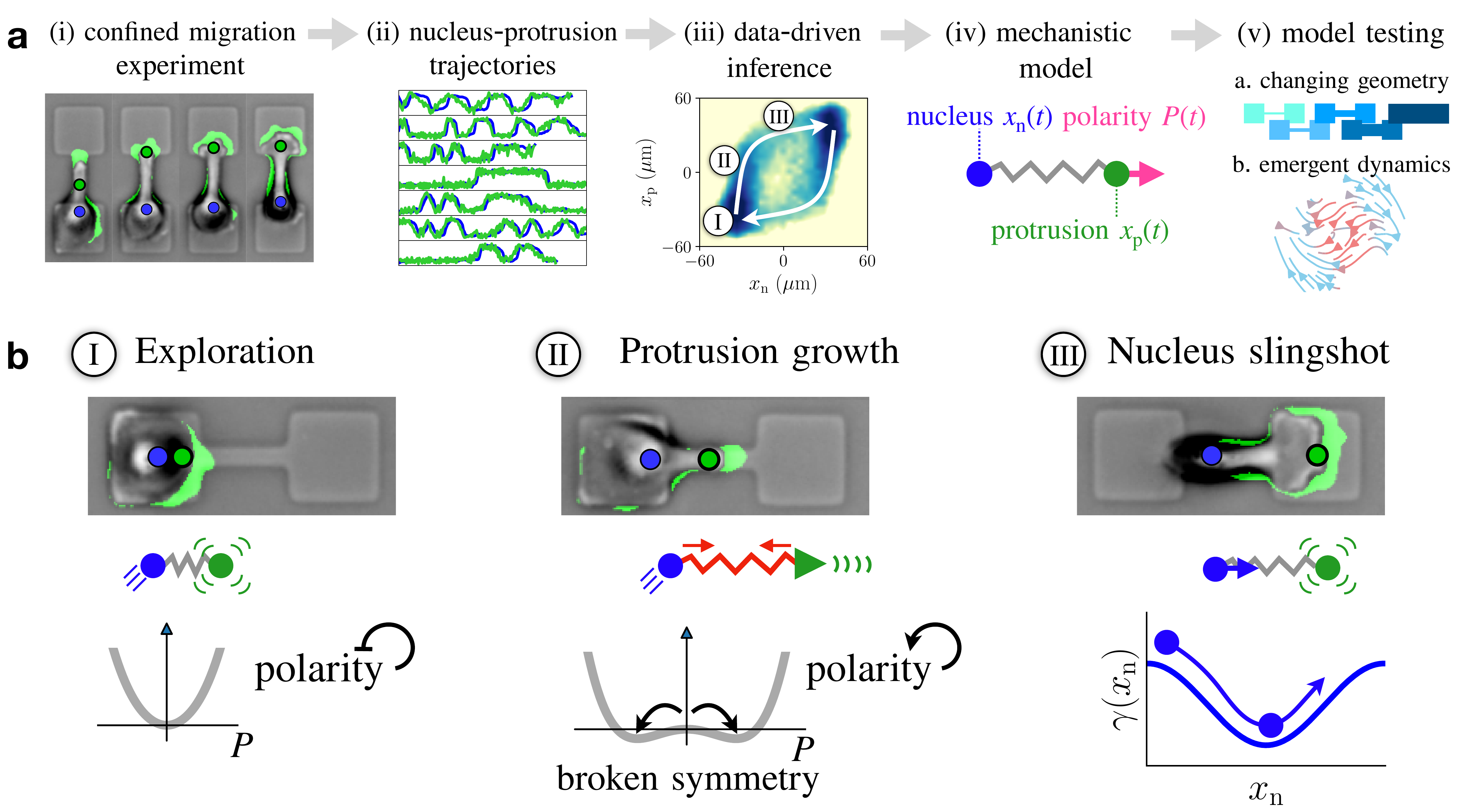}
	\centering
		\caption{
		\textbf{Data-driven development of a mechanistic model for confined cell migration.}
		\textbf{a}, Schematic of our approach. We generate a large data-set of joint protrusion-nucleus trajectories in a confined migration experiment (i, ii), and develop a data-driven approach by inferring model terms in the phase-space of protrusion and nucleus positions (iii). White arrows in (iii) indicate the stereotypical protrusion-nucleus cycling observed experimentally. Based on this inference, we systematically constrain a mechanistic model for the coupled dynamics of cell nucleus, protrusion, and the polarity driving force acting on the protrusion (iv). We test this model on a set of independent experiments, and by predicting the emergent long time-scale dynamics of the system (v). 
		\textbf{b}, The model identifies three stages of protrusion-nucleus cycling. In the joint position probability distribution of nucleus $x_\mathrm{n}$ and protrusion $x_\mathrm{p}$ (panel \textbf{a},iii, same color code as in Fig.~\ref{fig1}f), we indicate the typical evolution of the system with white arrows, and identify three distinct stages of the process. Typical brightfield microscopy images with overlayed protrusive areas, and the positions $x_\mathrm{n}$ and $x_\mathrm{p}$, are shown for each of the three stages. Schematics indicate the physical mechanisms that determine each phase according to our model.
		 }
	\label{fig_sum}
\end{figure*}
%%%%%%%%%%%%%%%%%%%%%%%%%%%%%%%%%%%%%

The ability of cells to migrate is essential for many physiological processes, including embryogenesis, immune response, and cancer~\cite{Franz2002,Scarpa2016,Luster2005a,Friedl2003}. In all these processes, cell migration relies on the interplay of several cellular components, including the formation of cell protrusions~\cite{Pollard2003a,Caswell2018}, adhesive connections to the environment~\cite{Yamada1997,Cukierman2001}, and the positioning of the cell nucleus~\cite{Denais2016,Davidson2020,Davidson2021}. These components are coupled by the polarizable active cytoskeleton, and together play the dual role of sensing the cell's local microenvironment and driving its net motion. At the cellular scale, this machinery leads to coordinated, functional migration, which manifests as persistent random motion on uniform two-dimensional substrates~\cite{Gail1970,Selmeczi2005}. However, in physiologically relevant contexts, cells must navigate complex, structured extracellular environments~\cite{Petrie2012,Caswell2018}, featuring obstacles such as thin constrictions~\cite{Friedl2003,Paul2017}. Thus, migrating cells may adapt their migration strategy, and the underlying  protrusion and polarity dynamics,  by responding to the structure of their local micro-environment. 

At the scale of whole-cell trajectories, confined cells exhibit intricate stochastic nonlinear dynamics in position-velocity phase space, such as limit cycles and bistability~\cite{Brueckner2019}. These findings and other studies~\cite{Paul2016a,Mahmud2009,Caballero2014,Caballero2015,LoVecchio2020,Fink2019,Brueckner2020,Brueckner2020a,Metzner2015,Ron2020a,Davidson2020,Reversat2020,Renkawitz2019} indicate that the migratory dynamics of cells are strongly affected by the presence of a confinement. However, the underlying physical principles and mechanisms that determine these dynamics remain elusive. Specifically, it remains unclear if the dynamics of cells actively adapt to external confinement, or whether confinements simply serve as passive boundaries. The search for such adaptive mechanisms is complicated by the intertwined behavior of the various cellular components and features, such as cell shape, protrusions, polarity, and nucleus, which could factor into this problem. Achieving a mechanistic understanding of protrusion and polarity dynamics in confined cell migration could yield key insights into both the underlying molecular mechanisms and the biological functions associated with these dynamics. 

To connect underlying mechanisms to the emergent behavior of migrating cells, bottom-up mechanistic approaches are a promising avenue. These include complex computational models modelling polarity processes and protrusion formation, including phase-field~\cite{Shao2010,Shao2012,Ziebert2012a} and Cellular Potts models~\cite{Graner1992,Segerer2015,Thuroff2019a,Goychuk2018}. More coarse-grained models include active particle models~\cite{Romanczuk2012a}, active gel theories~\cite{Kruse2006,Recho2019}, molecular clutch models~\cite{Chan2008,Elosegui-Artola2018}, and models coupling actin flow, polarity cues, and focal adhesion dynamics~\cite{Maiuri2015,CallanJones2016,Ron2020a,Sens2020,Hennig2020,Schreiber2021}. An orthogonal avenue to these bottom-up models are top-down approaches that infer cellular dynamics directly from observed trajectories~\cite{Selmeczi2005,Metzner2015,Brueckner2019,Brueckner2021,LaChance2022}. However, a direct connection of mechanistic bottom-up models to data-driven top-down perspective has remained difficult due to two main reasons. First, mechanistic models often contain many parameters that are hard to constrain experimentally. Thus, a crucial challenge is to reduce a mechanistic description to a level that can be constrained by data, while still capturing key behaviors of the important cellular components. The second challenge is to obtain large experimentally measured trajectory data sets of cellular features that allow us to learn such a minimal mechanistic description.

Here, we develop a hybrid data-driven and mechanistic approach, where we use experimental data to systematically constrain a minimal mechanistic model for confined cell migration postulated on the basis of physical principles and known cellular processes. To constrain this model, we experimentally study cells confined to a controlled micropatterned environment, allowing us to systematically vary the degree of confinement (Fig.~\ref{fig_sum}a,i). By observing the cell shapes in these experiments, we generate a large data set of joint nucleus and protrusion trajectories (Fig.~\ref{fig_sum}a,ii). Interestingly, under strong confinement, we find that cells exhibit a stereotypical migration pattern, which we term `protrusion-nucleus cycling' (Fig.~\ref{fig_sum}a,iii). Using a data-driven approach, we constrain a mechanistic description of the nucleus and protrusion dynamics by systematically increasing model complexity (Fig.~\ref{fig_sum}a,iv). This approach reveals two key insights into the confined migration dynamics: first, we find that the average dynamics of the nucleus are determined by an adhesion landscape describing the locally available adhesive area. Second, the cell polarity, which drives the protrusions, couples to the local confining geometry by switching from a negative to a positive, self-reinforcing feedback loop under strong confinement. Importantly, this mechanistic model accurately predicts cellular dynamics in systems with varying constriction width and length (Fig.~\ref{fig_sum}a,v). Thus, by systematically disentangling the contributions of nucleus, protrusions, and polarity to the cellular dynamics we identify a mechanism of polarity adaptation to confinements which plays a key role in the behavioral dynamics of confined cells.

%%%%%%%%%%%%%%%%%%%%%%%%%%%%%%%%%%%%%
%%FIGURE 
\begin{figure*}[ht]
	\includegraphics[width=0.8\textwidth]{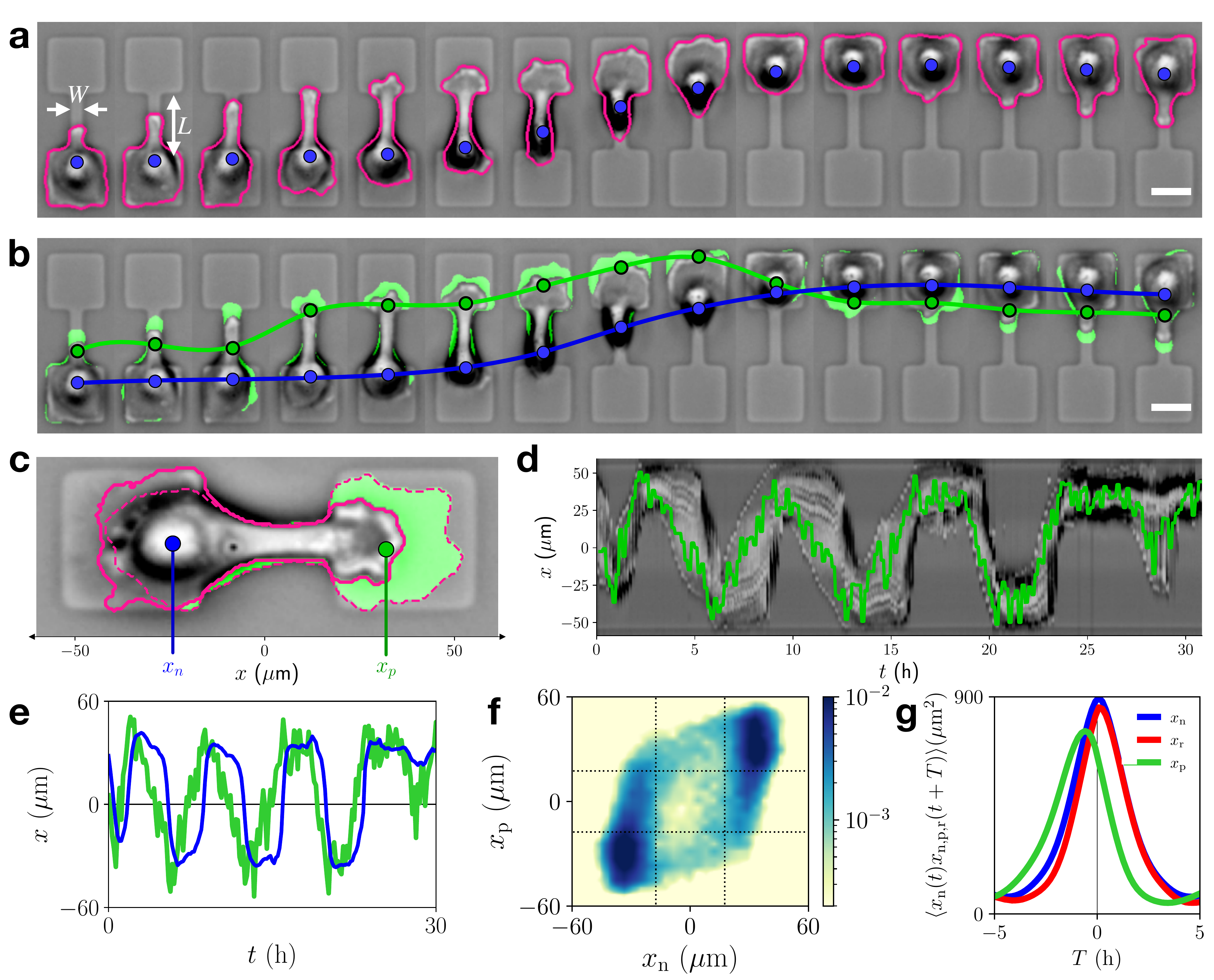}
	\centering
		\caption{
		\textbf{Extracting protrusion-nucleus dynamics from confined migration experiments.}
		\textbf{a.} Exemplary brightfield microscopy image series of an MDA-MB-231 breast cancer cell migrating in a two-state micropattern with constriction width $W=7 \ \si{\micro\meter}$ and length $L=35 \ \si{\micro\meter}$, indicated by arrows. Images are inverted for better visibility of cell shapes. Tracked cell shapes are shown as pink outlines. Blue dots indicate tracked nucleus position $x_\mathrm{n}$.
		\textbf{b.} Same time-series as in \textbf{a}, with protrusive areas marked in green, blue dot indicates the nucleus position $x_\mathrm{n}$, green dot the effective protrusion position $x_\mathrm{p}$. Green and blue lines indicate the trajectories of these two coordinates.
		\textbf{c.} Example image showing how protrusion areas are calculated. The solid pink line shows the current boundary of the cell area at time $t$, and the dashed line is the boundary at $t+\Delta t$. Protrusive area is shown in green.
		\textbf{d.} Kymograph of the brightfield microscopy images, with superimposed protrusion trajectory $x_\mathrm{p}(t)$ in green.
		\textbf{e.} Joint trajectory of nucleus $x_\mathrm{n}(t)$ (blue) and protrusion $x_\mathrm{p}(t)$ (green).
		\textbf{f.} Joint probability distribution $p(x_\mathrm{n},x_\mathrm{p})$ of the $x$-positions of nucleus and protrusion, plotted logarithmically and interpolated. Dotted lines indicate the boundaries of the adhesive islands.
		\textbf{g.} Position cross-correlations between nucleus and protrusion $\langle x_\mathrm{n}(t) x_\mathrm{p}(t+T)\rangle$ (green), and nucleus and retraction $\langle x_\mathrm{n}(t) x_\mathrm{r}(t+T)\rangle$ (red). The nucleus-protrusion correlation exhibits a peak at negative time shifts, indicating that the protrusion leads the nucleus by a typical time-shift $T_\mathrm{np} \approx 0.6 \ \mathrm{h}$. The retraction position $x_\mathrm{r}$ is determined in a similar way to the protrusion (Appendix~\ref{sec_image}). Blue line shows the nucleus position auto-correlation, $\langle x_\mathrm{n}(t) x_\mathrm{n}(t+T)\rangle$.
		All scale bars: $25  \ \si{\micro\meter}$.
				 }
	\label{fig1}
\end{figure*}
%%%%%%%%%%%%%%%%%%%%%%%%%%%%%%%%%%%%%

\section{Results}
%------------------------------------------------------------
\subsection*{Protrusion dynamics drive confined cell migration}

To investigate the dynamics of cell shapes, protrusions, and nucleus in confined migration, we study the migration dynamics of single MDA-MB-231 breast carcinoma cells confined to two-state micropatterns (Fig.~\ref{fig1}a). These patterns consist of two adhesive islands connected by a thin adhesive bridge, allowing us to study how migrating cells respond to constrictions in the extra-cellular environment. We use time-lapse phase-contrast microscopy and fluorescent staining of the cell nuclei to investigate the joint dynamics of cell shape and nucleus motion. We find that the motion of the nucleus is correlated with the growth of a protrusion across the constriction of the pattern, suggesting that the protrusion dynamics of these cells are key to understanding cell migration dynamics. 

To quantify these protrusive dynamics, we first isolate cell shapes from bright-field microscopy image stacks using a convolutional neural network with a U-Net architecture~\cite{Ronneberger2015} (Methods). This segmentation procedure allows us to accurately determine the 2D shape of the cells as a function of time (Fig.~\ref{fig1}a). To identify protrusions, we classify those components of the cell shape added in each time step as protrusive areas (Fig.~\ref{fig1}b,c, Methods, Supplementary Movie S3) \cite{Machacek2006}. During the traversal of cells across the constriction, large protrusive areas are formed at the leading edge of the cell. Importantly, due to the micropattern geometry, most protrusive activity is in the $x$-direction along the long axis of the pattern (Appendix~\ref{sec_image}). Thus, to provide a low-dimensional representation of the protrusion dynamics, we define the effective protrusion position $x_\mathrm{p}(t)$ as the $x$-component of the geometric center of protrusive area (Fig.~\ref{fig1}c), referred to as the protrusion from here on. Indeed, the protrusion trajectories serve as an indicator of the protrusive dynamics of the cells, as shown by an overlay with the kymograph of the microscopy images (Fig.~\ref{fig1}d, Supplementary Movie S3). In addition, we track the trajectories of the cell nucleus. While the cells also perform retractions at the trailing edge, we find that these are strongly correlated with the motion of the nucleus with near-zero time-lag, and therefore do not contain significant additional information (Fig.~\ref{fig1}g, Appendix~\ref{sec_image}). Thus, we  restrict our analysis to the nucleus-protrusion dynamics. This analysis pipeline gives access to a large data set of low-dimensional trajectories of cell nucleus and protrusion dynamics (1400 trajectory pairs with duration up to 50h), allowing an in-depth statistical analysis of the cellular dynamics.

The joint nucleus and protrusion trajectories reveal that these cells tend to migrate across the constriction in a stereotypical manner: first, the protrusion grows slowly across the constriction, after which the nucleus rapidly follows (Fig.~\ref{fig1}e). The nucleus motion exhibits weaker fluctuations than the protrusions and responds to the protrusions with a time delay, as quantified by the cross-correlation function $\langle x_\mathrm{n}(t) x_\mathrm{p}(t+T) \rangle$ (Fig.~\ref{fig1}g). The stereotypical migration pattern is reflected as a ring-like structure in the joint probability distribution of nucleus and protrusion positions $p(x_\mathrm{n},x_\mathrm{p})$  (Fig.~\ref{fig1}f). While the most likely states are where both nucleus and protrusion occupy one island, there is significant probability along the path where the protrusion first crosses the constriction and reaches the other island, followed by the traversal of the nucleus. In contrast, there is  low probability of observing both protrusion and nucleus in the constriction. Together, these results indicate that the confined migration dynamics exhibit a stereotypical `protrusion-nucleus cycling' represented as paths in $x_\mathrm{n}x_\mathrm{p}$-space. 

%%%%%%%%%%%%%%%%%%%%%%%%%%%%%%%%%%%%%
%%FIGURE 
\begin{figure}[ht]
	\includegraphics[width=0.5\textwidth]{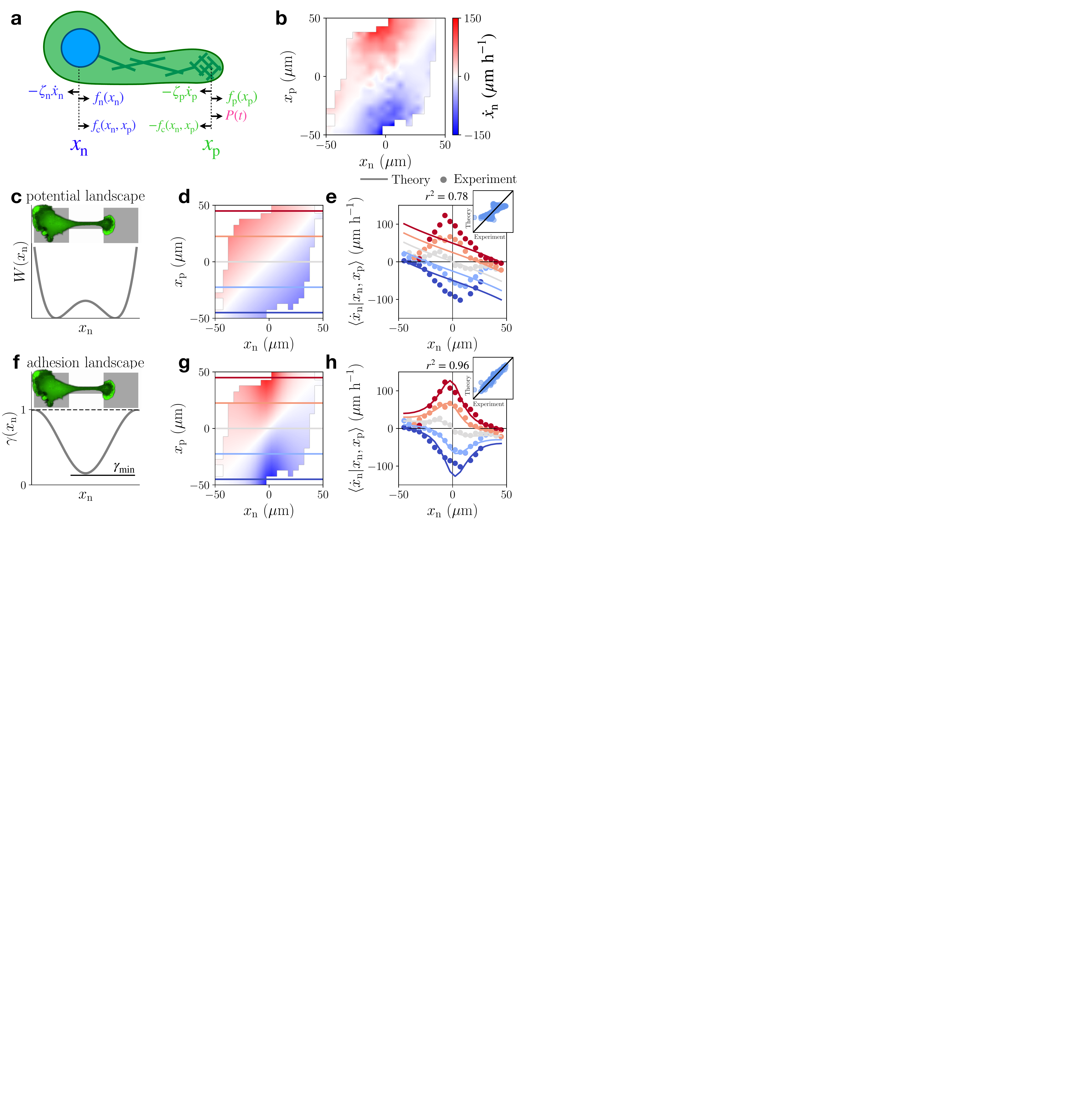}
	\centering
		\caption{
		\textbf{Nucleus velocity maps constrain model candidates.} 
		\textbf{a.} Schematic of the model. Arrows indicate the forces acting on the cell at positions $x_\mathrm{n}$ and $x_\mathrm{p}$.
		\textbf{b.} Experimental nucleus velocity map (NVM), calculated as the conditional average of the nucleus velocity as a function of nucleus and protrusion positions, $\langle \dot{x}_\mathrm{n} | x_\mathrm{n},x_\mathrm{p} \rangle$, shown with interpolation.
		\textbf{c.} Double-well potential $W(x_\mathrm{n})=Q(1-(x_\mathrm{n}/x_0)^2)^2$, where $Q$ determines the height of the energy barrier, and $x_0$ the positions of the minima. Image indicates the dimensions of the micropattern, and shows a fluorescence microscopy image of the actin cytoskeleton of a confined cell (LifeAct-GFP-transfection).
		\textbf{d.} NVM predicted by the energy potential. Parameters are determined by a best fit to the experimental NVM with two free fit parameters $k/\zeta_n$ and $Q$ (Appendix~\ref{sec_implementation}).
		\textbf{e.} Cuts of the NVM along the horizontal lines indicated in panel \textbf{d}, showing $\dot{x}_\mathrm{p}$ as a function of $x_\mathrm{n}$ for different $x_\mathrm{p}$. Dots: Experiment, solid lines: deformation model prediction.
		\textbf{f.} Spatially variable friction $\gamma(x_\mathrm{n}) = \gamma_\mathrm{min} + \frac{1}{2}(1-\gamma_\mathrm{min}) \left(1 - \cos \left(x_\mathrm{n} \pi / L_\mathrm{system} \right) \right)$ used in the adhesion landscape model, where $\gamma_\mathrm{min}$ is the friction at the center of the constriction, and $L_\mathrm{system}$ is the total length of the micropattern.
		\textbf{g.} NVM predicted by the adhesion landscape model with two free fit parameters $k/\zeta_n$ and $\gamma_\mathrm{min}$.  		
		\textbf{h.} Same plot as in panel \textbf{e} for the adhesion landscape model. 
						 }
	\label{fig2}
\end{figure}
%%%%%%%%%%%%%%%%%%%%%%%%%%%%%%%%%%%%%

%------------------------------------------------------------
\subsection*{Confined cells migrate in an adhesion landscape}

We aim to develop a mechanistic theory to describe how the coupled stochastic dynamics of cell nucleus and protrusion determine the confined migration of cells. Our strategy will be to postulate simple model candidates based on known cellular processes, physical principles, and symmetry arguments, which we systematically and quantitatively validate with experimental data. This approach allows us to rule out a whole range of possible alternative models, and identify a promising mechanistic model with predictive power for the protrusion-nucleus dynamics of confined cells.

Experimentally, we find that fluctuations in the nucleus velocities $\dot{x}_\mathrm{n}$ are small compared to the average components. By comparison, the fluctuations in the protrusion velocities $\dot{x}_\mathrm{p}$ are much larger and dominate over deterministic contributions to the protrusion velocities (Appendix~\ref{sec_whitenoise}). Thus, we consider a model in which the intrinsic stochasticity of the system stems from the polarity dynamics driving the protrusion. We expect forces on the nucleus to arise due to two main contributions: coupling to the cell protrusion~\cite{Crisp2006,Caswell2018,Davidson2020,Davidson2021}, and the effect of the confining micropattern. Similarly, protrusions couple to the cell nucleus~\cite{Caswell2018}, and may be sensitive to the external environment. Taken together, considering force balance at $x_\mathrm{n}$ and $x_\mathrm{p}$ (Fig.~\ref{fig2}a), we obtain 
\begin{align}
\label{eqn_eom_n}
\zeta_\mathrm{n} \dot{x}_\mathrm{n} &= f_\mathrm{c}(x_\mathrm{n},x_\mathrm{p}) + f_\mathrm{n}(x_\mathrm{n})  \\
\label{eqn_eom_p}
\zeta_\mathrm{p} \dot{x}_\mathrm{p} &= -f_\mathrm{c}(x_\mathrm{n},x_\mathrm{p}) + f_\mathrm{p}(x_\mathrm{p}) + P(t)
\end{align}
where $\zeta_\mathrm{n}, \zeta_\mathrm{p}$ are the friction coefficients of nucleus and protrusion, respectively, $f_\mathrm{c}$ is the coupling between nucleus and protrusion, and $f_\mathrm{n,p}$ are additional forces acting on each nucleus and protrusion due to the confinement. Additionally, we assume the protrusion to be driven by a stochastic active force $P(t)$, which serves as a minimal implementation of the time-dependent forces driving protrusion formation, such as the active pushing force due to actin polymerization~\cite{Pollard2003a,CallanJones2016}. This active force determines the instantaneous direction of polarization in which protrusions are generated, and we therefore refer to it as the cell polarity. 

To constrain our model step-by-step, we start with the dynamics of the cell nucleus. In migrating cells, the motion of the nucleus is coupled to the dynamics of the leading edge, for example, through material stresses in the cytoskeleton connecting protrusion and nucleus~\cite{Crisp2006,Caswell2018,Davidson2020a,Davidson2020}, or through mechanical feedback processes coupling the leading and trailing edge of the cell~\cite{Tsai2019}. As a minimal model for this coupling, we consider a linear elastic spring, similar to previous work~\cite{Ron2020a,Sens2020}.

It is less clear, however, how to incorporate the effect of the confining micropattern on the dynamics. Physically, we consider two distinct ways to couple the cell dynamics to geometry. First, a conservative force, corresponding to a double-well potential $W(x_\mathrm{n})$, with minima on the adhesive islands and a barrier around the constriction (Fig.~\ref{fig2} c), can provide a model for the contribution due to cell deformations during the transition. Such deformation dynamics of cells are frequently modelled using effective Hamiltonians including the surface and line tension of the cell~\cite{Albert2014,Bi2014,Bi2015,Segerer2015,Goychuk2018}, which would suggest that the deformed state of the cell in the constriction is associated with an increased energy. Second, the difference in adhesive area available to the cell on the island and in the constriction could lead to a dissipative force corresponding to a spatially variable friction coefficient~\cite{Tawada1991,Reboux2008}. Mesenchymal migration exhibits mature focal adhesions at the cell rear, where the nucleus typically resides~\cite{Lehnert2004a,Balaban2001}. These adhesions can only form within the micropatterned area, and we therefore expect the adhesiveness to be largest on the islands and smallest at the center of the constriction. 

%%%%%%%%%%%%%%%%%%%%%%%%%%%%%%%%%%%%%
%%FIGURE 
\begin{figure}[h]
	\includegraphics[width=0.45\textwidth]{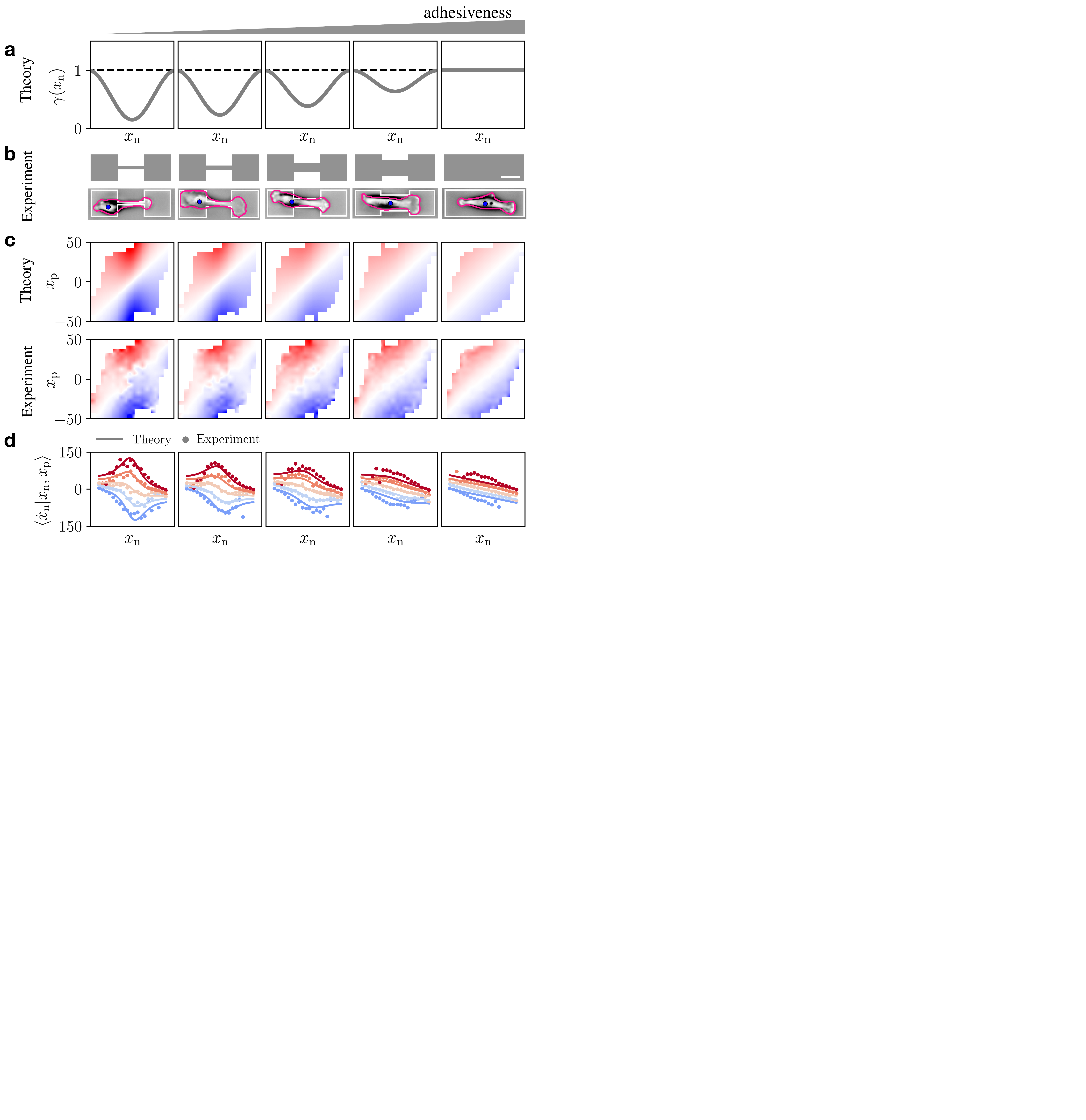}
	\centering
		\caption{
		\textbf{Adhesion landscape model predicts dynamics with varying constriction width.} 
		\textbf{a.} Friction profiles $\gamma(x_\mathrm{n})$ with increasing $\gamma_\mathrm{min}$ as a model for increasing bridge width. The value of $\gamma_\mathrm{min}$ is fitted for the narrowest bridge ($W=4 \ \si{\micro\meter}$; $\gamma_\mathrm{min} \approx 0.2, k/\zeta_n \approx 0.6$ h$^{-1}$). For the widest system without constriction, we take a flat profile and take intermediate values of $\gamma_\mathrm{min}$ for intermediate widths, such that $\gamma_\mathrm{min}$ scales linearly with the available area (Appendix~\ref{sec_implementation}).
		\textbf{b.} Sketch of confinement geometries with increasing bridge widths $W=4,7,12,22,35 \ \si{\micro\meter}$ (from left to right); brightfield microscopy images of MDA-MB-231 cells migrating in these geometries with cell outline in pink and nucleus position in blue, and geometry in white. Scale bar: $25  \ \si{\micro\meter}$.
		\textbf{c.} Predicted and experimental nucleus velocity maps (NVM) $\langle \dot{x}_\mathrm{n} | x_\mathrm{n},x_\mathrm{p} \rangle$. Plotted with the same color axis as shown in Fig.~\ref{fig2}.
		\textbf{d.} Predicted and experimental cuts of the NVM, showing $\dot{x}_\mathrm{p}$ as a function of $x_\mathrm{n}$ for different $x_\mathrm{p}$ as described in Fig.~\ref{fig2}. 
		 }
	\label{fig3}
\end{figure}
%%%%%%%%%%%%%%%%%%%%%%%%%%%%%%%%%%%%%

To test the energy potential contribution, we consider the equation of motion for the cell nucleus
\begin{align}
\label{eqn_doublewell}
\zeta_\mathrm{n} \dot{x}_\mathrm{n} = k(x_\mathrm{p}-x_\mathrm{n}) - \partial_{x_\mathrm{n}} W(x_\mathrm{n})
\end{align}
This equation makes a concrete prediction for how the nucleus velocity $\dot{x}_\mathrm{n}$ varies with the positions of nucleus and protrusion. To test this prediction directly on the experimental data, we determine the average velocity of the cell nucleus as a function of $x_\mathrm{n}$ and $x_\mathrm{p}$, $\langle \dot{x}_\mathrm{n} | x_\mathrm{n},x_\mathrm{p} \rangle$, which we term the \emph{nucleus velocity map} (NVM) (Fig.~\ref{fig2}b). Importantly, with this approach based purely on the nucleus velocities, we can determine the deterministic nucleus dynamics (Eq.~\eqref{eqn_eom_n}) without making  assumptions about the protrusion and polarity dynamics (Eq.~\eqref{eqn_eom_p}). However, we find that the NVM predicted by the energy potential fails to capture the experimental data, as it does not predict the characteristic acceleration of the nucleus in the constriction (Fig.~\ref{fig2}d,e). This approach similarly fails for more general non-linear elastic couplings between nucleus and protrusion (Appendix~\ref{sec_strategy}). Therefore, we conclude that such potential energy models alone are not able to recover the cellular dynamics in this setup.

To test the possible contribution of differences in local adhesion, a simple model is a spatially variable friction coefficient:
\begin{align}
\label{eqn_adhesion}
\zeta_\mathrm{n} \gamma(x_\mathrm{n}) \dot{x}_\mathrm{n} = k(x_\mathrm{p}-x_\mathrm{n})
\end{align}
where $\gamma(x_\mathrm{n})$ ensures lower friction in the constriction (Fig.~\ref{fig2}f). This model provides an excellent fit to our data, and captures the characteristic increase in nucleus speeds during traversal (Fig.~\ref{fig2}g,h). The resulting fit parameters give a typical time-scale for the nucleus motion $\zeta_n/k \approx 1.7$h, which is in approximate agreement with known turn-over times of focal adhesions~\cite{Stricker2013a,Stehbens2014}. This time-scale is reduced in the constriction due to the reduced number of adhesions formed by the cell body around the nucleus when it is in the constriction, causing the acceleration of the nucleus during traversal. Taken together, these results indicate that a dissipative force arising from a spatially variable adhesion landscape is a key component of the effect of the confining constriction on migration dynamics, which in our setup appears to dominate over possible contributions due to cellular deformations.

%------------------------------------------------------------
\subsection*{Adhesion landscape model captures dependence of nucleus dynamics on constriction width}

The adhesion landscape model (Eq.~\eqref{eqn_adhesion}) makes a simple, intuitive prediction. As we widen the constricting bridge of the micropattern, more adhesive area becomes available, thereby reducing the variations in the friction profile (Fig.~\ref{fig3}a). In the  limit where the constriction has the same width as the islands, we expect a uniform adhesiveness profile. Accordingly, we predict the acceleration of the cell nucleus observed on thin bridges (Fig.~\ref{fig2}) to decrease with the increasing adhesiveness of a wider bridge, and to completely disappear for constant adhesiveness (Fig.~\ref{fig3}c,d). In this limiting case, we expect migration dynamics that are completely determined by the linear elastic coupling between nucleus and protrusion (last panel Fig.~\ref{fig3}d). 

To challenge the predictive power of the adhesion landscape model, we perform experiments with varying bridge width (Fig.~\ref{fig3}b). Importantly, the nucleus velocity maps inferred from these experiments are well predicted by the model, and exhibit the predicted decreasing maximum nucleus speed in the constriction (Fig.~\ref{fig3}c,d). On the rectangular micropattern without constriction, we find an almost linear profile of the nucleus speed with position, as predicted theoretically. This further supports our model of a linear elastic nucleus-protrusion coupling. In summary, the adhesion landscape model has predictive power for confining geometries with varying constriction width.

%%%%%%%%%%%%%%%%%%%%%%%%%%%%%%%%%%%%%
%%FIGURE 
\begin{figure}[h!]
	\includegraphics[width=0.45\textwidth]{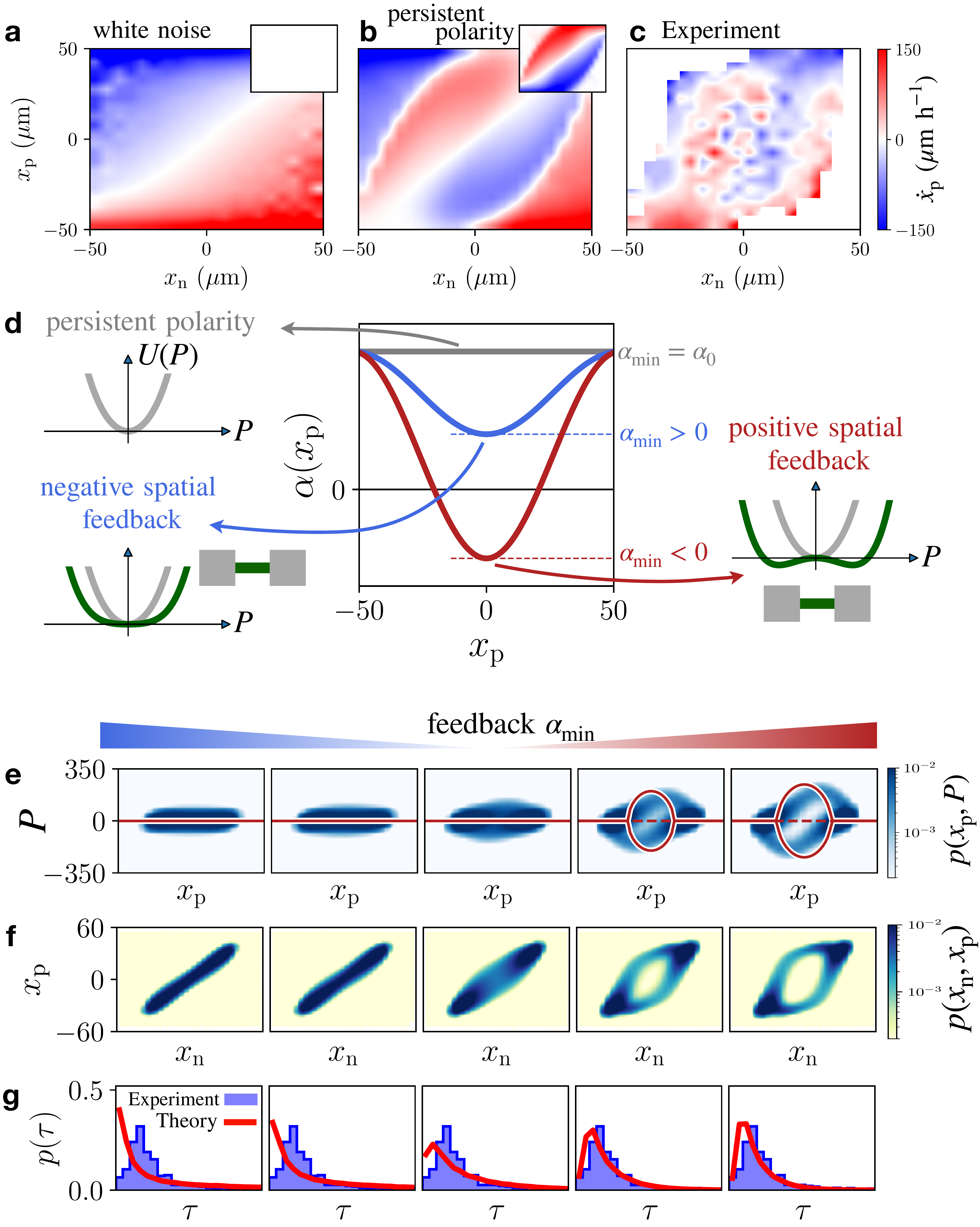}
	\centering
		\caption{
		\textbf{Protrusion velocity maps and the geometry adaptation model.} 
		\textbf{a,b.} Protrusion velocity maps (PVM) $\langle \dot{x}_\mathrm{p} | x_\mathrm{n},x_\mathrm{p} \rangle$ predicted by the white noise model, and the persistent polarity model. In both models, we use a potential to enforce the overall system boundaries, $V(x_\mathrm{p}) = (x/x_\mathrm{boundary})^8$ (Appendix~\ref{sec_implementation}).	 	
		\emph{Inset:} polarity contribution to the PVM, given by $\langle P | x_\mathrm{n},x_\mathrm{p} \rangle$.
		\textbf{c.}  PVM inferred from experiments with bridge width $W=7 \ \si{\micro\meter}$.
		\textbf{d.} Schematic illustration of the polarity models. The persistent polarity corresponds to a flat feedback profile, $\alpha(x_\mathrm{p})=\alpha_0$, where the polarity is effectively confined to a constant harmonic potential $U(P)$ (left). In the geometry adaptation model (Eq.~\eqref{eqn_Kxp}), $\alpha_\mathrm{min}$ controls the sign of the feedback. For $\alpha_\mathrm{min}>0$, the feedback is negative and the polarity is confined to harmonic potentials with spatially varying stiffness (left). For $\alpha_\mathrm{min}<0$, the feedback locally becomes positive, leading to two stable fixed points (right).
		\textbf{e.} From left to right, we vary the value of $\alpha_\mathrm{min}= \{6,4,0,-4,-6\} \ \mathrm{h}^{-1}$. The joint probability distribution of protrusion position and polarity $p(x_\mathrm{p},P)$ is shown. Solid red lines indicate the position of the stable fixed points in the polarity dynamics, dashed red lines indicate unstable fixed points.
		\textbf{f.} Joint probability distributions $p(x_\mathrm{n},x_\mathrm{p})$ predicted by the geometry adaptation model with varying $\alpha_\mathrm{min}$. 
		\textbf{g.} Probability distribution of the dwell times $\tau$ predicted by the model (red) and observed experimentally ($W=7 \ \si{\micro\meter}$, blue).
		 }
	\label{fig_Kxp}
\end{figure}
%%%%%%%%%%%%%%%%%%%%%%%%%%%%%%%%%%%%%

%------------------------------------------------------------
\subsection*{Protrusions driven by time-correlated polarity}

Having determined how the dynamics of the nucleus couples to the confinement and protrusion, we next investigate the dynamics of the protrusion itself (Eq.~\eqref{eqn_eom_p}). As a minimal model for the protrusion dynamics, we postulate a coupling to the cell nucleus equal and opposite to the coupling introduced for the cell nucleus (Eq.~\eqref{eqn_adhesion}). In addition, we enforce a potential $V(x_\mathrm{p}) = (x/x_\mathrm{boundary})^8$ to confine the protrusion between the boundaries of the micropattern (Appendix~\ref{sec_implementation}):
\begin{align}
\label{eqn_protrusion}
\zeta_\mathrm{p} \dot{x}_\mathrm{p} = -k(x_\mathrm{p}-x_\mathrm{n}) - \partial_{x_\mathrm{p}} V(x_\mathrm{p}) + P(t)
\end{align}
Thus, we assume that both the friction and the potential term of the protrusion are insensitive to the presence of the constriction. Importantly, however, we anticipate that the polarity $P$ may couple to the confinement, as it models the active driving of the protrusion by the migration machinery, including actin polymerization and the diffusion of polarity cues~\cite{Pollard2003a,CallanJones2016}, which may be sensitive to the geometry of the confinement. 
%Furthermore, we also expect that the polarity may exhibit time-correlations to account for the spatiotemporal dynamics of the migration machinery.

Similar to our approach to the nucleus dynamics, this protrusion model provides a prediction for the average protrusion velocity as a function of $x_\mathrm{n}$ and $x_\mathrm{p}$, $\langle \dot{x}_\mathrm{p} | x_\mathrm{n},x_\mathrm{p} \rangle$, which we term \emph{protrusion velocity map} (PVM). According to our general model ansatz (Eq.~\eqref{eqn_eom_p}), unlike the NVM, the PVM consists of several components, including the polarity dynamics:
\begin{equation}
\label{eqn_PVM}
\langle \dot{x}_\mathrm{p} | x_\mathrm{n},x_\mathrm{p} \rangle = -f_\mathrm{c}(x_\mathrm{n},x_\mathrm{p}) + f_\mathrm{p}(x_\mathrm{p}) + \langle P | x_\mathrm{n},x_\mathrm{p} \rangle
\end{equation}
Since the polarity term $\langle P | x_\mathrm{n},x_\mathrm{p} \rangle$ does not average to zero for time-correlated polarities, we cannot in general disentangle the contributions to the protrusion dynamics based on the PVM~\cite{Lehle2018,Frishman2018}. Instead, we will constrain the polarity dynamics by systematically testing models of increasing complexity.

We first show that the data cannot be captured by the simplest possible stochastic polarity dynamics: a Gaussian white noise (WN) process $P_\mathrm{WN}= \sigma \xi(t)$ with $\langle \xi(t)\rangle=0$ and $\langle \xi(t)\xi(t') \rangle=\delta(t-t')$. In this case,  $\langle P_\mathrm{WN} | x_\mathrm{n},x_\mathrm{p} \rangle=0$ (Inset Fig.~\ref{fig_Kxp}a), and we directly recover the expected contractile elastic coupling (Eq.~\eqref{eqn_protrusion}) in the PVM (Fig.~\ref{fig_Kxp}a). In clear contrast to this prediction, the PVM inferred from experiments shows an intricate dependence of the protrusion velocities as a function of $x_\mathrm{n}$ and $x_\mathrm{p}$ (Fig.~\ref{fig_Kxp}c). These results indicate that to account for the experimentally observed dynamics, we need to account for time correlations in the polarity.

The overall structure of the experimental PVM is in line with a contractile coupling between nucleus and protrusion: it exhibits negative velocities for $x_\mathrm{p}>x_\mathrm{n}$ and positive velocities for $x_\mathrm{p}<x_\mathrm{n}$. These features correspond to the protrusion being pulled back towards the nucleus. However, when the protrusion extends into the constriction, the protrusion velocity switches sign, corresponding to an unexpected driving force pushing the protrusion away from the nucleus. This `polarity driving' cannot be accounted for even by a non-linear contractile coupling to the nucleus in our model with a white-noise polarity. Instead, we expect that the polarity may exhibit time-correlations to account for the spatiotemporal dynamics of the migration machinery. 

To investigate how time-correlated polarity dynamics affect the migration behavior, we consider the simplest choice of a persistent, exponentially correlated polarity 
\begin{equation}
\label{eqn_POU}
\dot{P}_\mathrm{per} = - \alpha_0 P_\mathrm{per} + \sigma \xi(t)
\end{equation}
with $\alpha_0>0$. In this case, the polarity experiences negative feedback, $\dot{P}_\mathrm{per} \propto -P_\mathrm{per}$, and exhibits time-correlations $\langle P_\mathrm{per}(0) P_\mathrm{per}(t) \rangle$ decaying exponentially on a persistence time-scale $\alpha_0^{-1}$. These polarity dynamics have significant correlations with the state of the system: $\langle P_\mathrm{per} | x_\mathrm{n},x_\mathrm{p} \rangle \neq 0$ (Inset Fig.~\ref{fig_Kxp}b), and thus, unlike in the white-noise case, the polarity contributes to the PVM (Eq.~\eqref{eqn_PVM}). Specifically, the persistent polarity exhibits a polarity driving similar to the experimental PVM (Fig.~\ref{fig_Kxp}b,c). Taken together, these results indicate that cell protrusions are driven by time-correlated polarity dynamics.

%------------------------------------------------------------
\subsection*{Confinement triggers polarity self-reinforcement}

While the persistent polarity (Eq.~\eqref{eqn_POU}) describes the qualitative features of the protrusion velocities (Fig.~\ref{fig_Kxp}b,c), it predicts stochastic dynamics that do not capture the key features of the experiment. Specifically, it fails to capture the stereotypical protrusion-nucleus cycling indicated by the ring structure in the experimental probability distribution $p(x_\mathrm{n},x_\mathrm{p})$ (Fig.~\ref{fig_Kdx_xpdx}a). The persistent polarity relies on several simplifying assumptions. Firstly, we assumed the polarity to be insensitive to the local confinement, as the polarity dynamics do not explicitly depend on the position of the protrusion. Secondly, we assumed negative feedback, $\dot{P}_\mathrm{per} \propto -P_\mathrm{per}$. This means that the polarity is effectively confined to a harmonic potential $U(P_\mathrm{per})= \alpha_0 P_\mathrm{per}^2 /2$, and thus always driven back towards zero (Fig.~\ref{fig_Kxp}d).

To relax these assumptions, we propose a geometry adaptation (GA) model, where the strength and sign of the polarity feedback depend on the local geometry of the confinement. Thus, the feedback may vary with the position of the protrusion, $\alpha=\alpha(x_\mathrm{p})$. Physically, we expect that the polarity may become more persistent when the protrusion is in the constriction. Such an increase in persistence could be due to increased alignment of actin fibers~\cite{SoaresESilva2011,Bonelli2016,Jiang2005,Prager-Khoutorsky2011}, or more stable polarity cue gradients~\cite{CallanJones2016,Maree2012a,Vasilevich2021} when the protrusion is confined to a narrow constriction. To ensure that the polarity remains bounded, we include the next-order term allowed by symmetry $-\beta P^3$, with $\beta>0$, and allowing $\alpha<0$ locally:
\begin{equation}
\label{eqn_Kxp}
\dot{P}_\mathrm{GA} = - \alpha(x_\mathrm{p}) P_\mathrm{GA} -\beta P_\mathrm{GA}^3  + \sigma \xi(t)
\end{equation}
To account for larger persistence in the constriction, we choose a feedback function $\alpha(x_\mathrm{p})$ with a minimal value $\alpha_\mathrm{min}$ in the center of the constriction (Fig.~\ref{fig_Kxp}d). If $\alpha_\mathrm{min}>0$, the polarity dynamics exhibits a stable fixed point at $P_\mathrm{GA}=0$ everywhere (Fig.~\ref{fig_Kxp}d). In contrast, if $\alpha_\mathrm{min}<0$, the polarity is still driven back to $P_\mathrm{GA}=0$ on the islands, but in the constriction, two stable fixed points $P_\mathrm{GA}^*= \pm \sqrt{|\alpha|/\beta}$ appear. Consequently, when the protrusion is in the constriction and the polarity is small ($|P_\mathrm{GA}|<|P_\mathrm{GA}^*|$), a positive feedback mechanism $\dot{P}_\mathrm{GA} \propto P_\mathrm{GA}$ is activated, leading to a self-reinforcement of the polarity in the current direction of polarization, breaking the symmetry of the dynamics. 

We now explore the predictions of the geometry adaptation model by varying $\alpha_\mathrm{min}$ (Fig.~\ref{fig_Kxp}e). As expected, for negative polarity feedback ($\alpha_\mathrm{min}>0$), we find a polarity distribution $p(x_\mathrm{p},P)$ with polarities centered around $P=0$ at all positions $x_\mathrm{p}$. In contrast, for positive feedback, polarities preferentially take finite values in the constriction, yielding a ring-structure in the polarity distribution $p(x_\mathrm{p},P)$ (Fig.~\ref{fig_Kxp}e). The model then predicts protrusion-nucleus cycling in $p(x_\mathrm{n},x_\mathrm{p})$ and a marked peak in the dwell time distribution, defined as the time between subsequent transitions of the cell nucleus across the bridge centre, both in line with the experiment (Fig.~\ref{fig_Kxp}f,g). 

Up to this order of description, there are two possible alternative models allowed by symmetry which couple the polarity to the state of the system. First, instead of coupling the polarity feedback to the geometry, the overall amplitude of the polarity could depend on position. Second, instead of a position-dependent feedback, one could consider feedback that depends on the cell's extension $|x_\mathrm{n}-x_\mathrm{p}|$. However, these alternatives fail to capture our experimental observations (Appendix~\ref{sec_strategy}). Taken together, these results suggest that to capture the stereotypical protrusion-nucleus cycling, we require a geometry-sensitive polarity feedback. 

%%%%%%%%%%%%%%%%%%%%%%%%%%%%%%%%%%%%%
%%FIGURE 
\begin{figure}[h!]
	\includegraphics[width=0.48\textwidth]{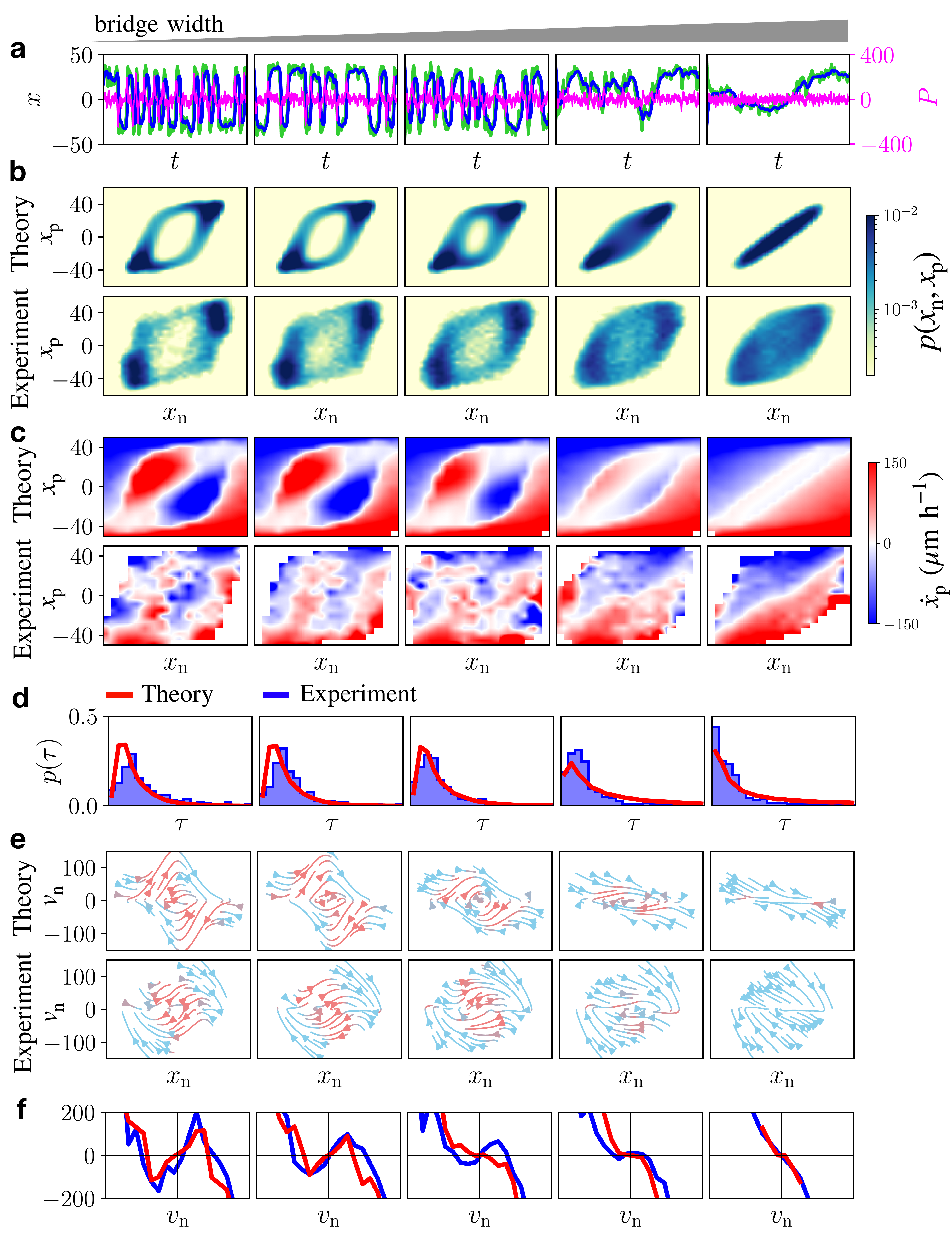}
	\centering
		\caption{
		\textbf{Geometry adaptation model predicts dynamics with varying constriction width.} 
		\textbf{a.}	Stochastic trajectories $x_\mathrm{n}(t)$ (blue), $x_\mathrm{p}(t)$ (green), and $P_\mathrm{GA}(t)$ (pink) predicted by our mechanistic model, which combines the adhesion landscape and geometry adaptation (Eq.~\eqref{eqn_adhesion},~\eqref{eqn_protrusion},~\eqref{eqn_Kxp}). 
		To predict dynamics with increasing bridge width, we simultaneously increase $\gamma_\mathrm{min}$ and $\alpha_\mathrm{min}$, while leaving all other parameters fixed (Appendix~\ref{sec_implementation}, Table~\ref{tab_params}). 
		\textbf{b.} Joint probability distributions $p(x_\mathrm{n},x_\mathrm{p})$. 
		\textbf{c.} Protrusion velocity maps (PVM) $\langle \dot{x}_\mathrm{p} | x_\mathrm{n},x_\mathrm{p} \rangle$. The top row corresponds to the model prediction, the bottom row to experimental observations.
		\textbf{d.} Predicted (red) and experimental (blue) dwell time distributions $p(\tau)$.
		\textbf{e.} Flow field $(\dot{x}_\mathrm{n},\dot{v}_\mathrm{n})=(v_\mathrm{n},F(x_\mathrm{n},v_\mathrm{n}))$ indicated by arrows~\cite{Brueckner2019}. Arrow color indicates the direction of the local flow: acceleration is orange and deceleration is blue.
		\textbf{f.} Predicted (red) and experimental (blue) effective friction at the bridge center $F(x_\mathrm{n} \to 0,v_\mathrm{n})$.
		In all panels, experimental observations correspond to $W=4,7,12,22,35 \ \si{\micro\meter}$ (from left to right).
		}
	\label{fig_w}
\end{figure}
%%%%%%%%%%%%%%%%%%%%%%%%%%%%%%%%%%%%%

%%%%%%%%%%%%%%%%%%%%%%%%%%%%%%%%%%%%%
%%FIGURE 
\begin{figure*}
	\includegraphics[width=\textwidth]{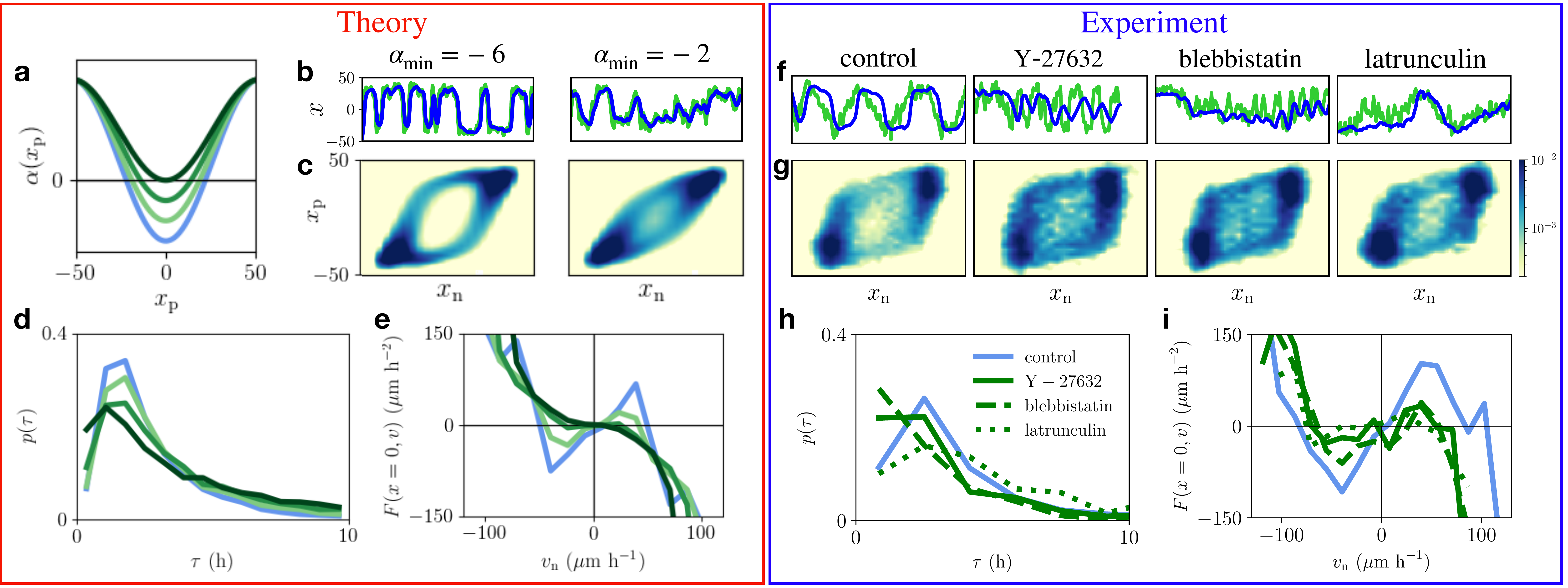}
	\centering
		\caption{
		\textbf{Geometry adaptation model captures response to pharmacological perturbation of polarity, contractility, and actin polymerization.} 
		\textbf{a.} Feedback profiles $\alpha(x_\mathrm{p})$ for varying strengths of the positive feedback, showing $\alpha_\mathrm{min}=\{-6,-4,-2,0\}$. 
		\textbf{b.} Model trajectories of $x_\mathrm{n}(t),x_\mathrm{p}(t)$ for the standard value $\alpha_\mathrm{min}=-6$, corresponding to blue curves in panels a,d,e; and $\alpha_\mathrm{min}=-2$, which provides a model for reduced polarity feedback. The adhesion profile is held constant, corresponding to no changes in the confinement geometry.
		\textbf{c.} Corresponding probability distributions $p(x_\mathrm{n},x_\mathrm{p})$.
		\textbf{d.} Dwell time distributions $p(\tau)$.
		\textbf{e.} Effective friction at the bridge center $F(x_\mathrm{n} \to 0,v_\mathrm{n})$.
		\textbf{f.} Experimental trajectories of $x_\mathrm{n}(t),x_\mathrm{p}(t)$ on micropatterns with constriction width $W=7\si{\micro\meter}$, for the control condition, treatment with the ROCK-inhibitor Y-27632, blebbistatin, and latrunculin. 		
		\textbf{g.} Corresponding probability distributions $p(x_\mathrm{n},x_\mathrm{p})$.
		\textbf{h.} Dwell time distributions $p(\tau)$.
		\textbf{i.} Effective friction at the bridge center $F(x_\mathrm{n} \to 0,v_\mathrm{n})$.
		}
	\label{fig_drugs}
\end{figure*}
%%%%%%%%%%%%%%%%%%%%%%%%%%%%%%%%%%%%%
 
%------------------------------------------------------------
\subsection*{Geometry adaptation model correctly predicts response to varying constriction dimensions}

Having constrained the model based on a single confining geometry, we challenge the predictive power of our approach by investigating the full stochastic trajectory dynamics of cells in micropatterns with varying constriction geometry. Specifically, we have fully constrained the model parameters for the adhesion landscape model for the nucleus (Eq.~\eqref{eqn_adhesion}; Fig.~\ref{fig2}) and the geometry adaptation model for protrusion and polarity (Eq.~\eqref{eqn_protrusion},~\eqref{eqn_Kxp}; Fig.~\ref{fig_Kxp}). Increasing constriction width has a clear implication for this model: in addition to a flattening adhesiveness profile $\gamma(x_\mathrm{n})$ (Fig.~\ref{fig3}a), we also expect the positive polarity feedback to diminish. The model predicts that the protrusion-nucleus cycling should disappear with increasing constriction width: the ring-structure of the position distribution $p(x_\mathrm{n},x_\mathrm{p})$ gradually closes (Fig.~\ref{fig_w}b), and the typical transition time-scale, indicated by the peak in the dwell time distribution, disappears (Fig.~\ref{fig_w}d). All these predicted features of the dynamics are quantitatively confirmed experimentally (Fig.~\ref{fig_w}b-d). Importantly, these predictions involve no further fitting: the key parameters $\{ k_\mathrm{n}, \gamma_\mathrm{min}, k_\mathrm{p}, \alpha_0, \alpha_\mathrm{min} \}$ for a thin constriction are completely determined in Fig.~\ref{fig2}, \ref{fig_Kxp}, and the variation of $\gamma_\mathrm{min},\alpha_\mathrm{min}$ with increasing width is determined by the micropattern geometry (Appendix~\ref{sec_implementation}, Table~\ref{tab_params}). 

The predictions of the model can be understood by examining the protrusion dynamics in the model: we find that the predicted PVM exhibits a polarity driving of decreasing magnitude with increasing bridge width (Fig.~\ref{fig_w}c). Notably, the driving disappears in the widest system with no constriction, where the protrusion velocities are thus determined by the protrusion-nucleus coupling. This coupling pulls protrusion and nucleus together, inhibiting the stereotypical cycle. The model therefore suggests that these stereotypical cycles rely on the adaptation of the cell polarity dynamics to its local environment. 

Furthermore, our mechanistic model explains the nonlinear dynamics of the cell nucleus motion, which can be described by an underdamped stochastic equation of motion: $\dot{v}_\mathrm{n} = F(x_\mathrm{n},v_\mathrm{n}) + \sigma(x_\mathrm{n},v_\mathrm{n}) \eta(t)$ where $\eta(t)$ is Gaussian white noise~\cite{Brueckner2019}. This underdamped equation of motion represents an effective description of the cellular dynamics, with no direct connection to cellular degrees of freedom such as the protrusion and polarity, which we consider here. The deterministic contribution $F(x_\mathrm{n},v_\mathrm{n})$ exhibits intricate non-linear dynamics, depicted in a phase-space portrait (Fig.~\ref{fig_w}e). These phase-space portrait reveal that the nucleus deterministically accelerates into the thin constriction (orange arrows Fig.~\ref{fig_w}e) which manifests as an effective `negative friction' at the center of the constriction (Fig.~\ref{fig_w}f). Our mechanistic model correctly predicts non-linear nucleus dynamics (Fig.~\ref{fig_w}e,f), and reveals that the observed deterministic acceleration is a consequence of two combined effects: lower adhesiveness and enhanced polarity persistence in the constriction (Appendix~\ref{sec_xv}).

We further challenge our model by exploring geometries with varying constriction length $L$, and the inclusion of multiple adhesive islands to create multi-state micropatterns (Supplementary Movie S8). We find that the model also captures the changes in dynamics observed in these systems (Appendix~\ref{sec_geom}). Taken together, these results indicate that our mechanistic model has predictive power beyond the specific confinement geometry which we used to constrain it.

%------------------------------------------------------------
\subsection*{Geometry adaptation depends on cell polarity, contractility and actin polymerization}

Key processes that affect the state of cell polarization are the structure and contractility of the actin network~\cite{Jiang2005,Prager-Khoutorsky2011,Trichet2012,Gupta2015,Ladoux2016}, as well as diffusable polarity cues~\cite{CallanJones2016,Hodge2016,Warner2019}, including Rho GTPase. To test whether geometry adaptation depends on these cellular components, we pharmacologically interfered with cell polarity, contractility, and protrusion formation. Specifically, we use inhibitors of Rho-associated protein kinase (ROCK) (Y-27632), myosin-II (blebbistatin) and actin polymerization (latrunculin) (Fig.~\ref{fig_drugs}f, Supplementary Movies S9-11). Interestingly, in the case of ROCK and myosin inhibition, we observe increased probability in the center of the ring-like probability distribution $p(x_\mathrm{n},x_\mathrm{p})$, a disappearance of the peak in the dwell time distribution, and a reduction of the negative friction in the nonlinear nucleus dynamics (Fig.~\ref{fig_drugs}g-i). The effect of actin polymerization inhibition are less pronounced, but qualitatively similar. This set of observed changes to the dynamics are in congruence with the predictions of our model following a reduction of the feedback strength $\alpha_\mathrm{min}$ (Fig.~\ref{fig_drugs}a-e). Importantly, changing other aspects of the model, such as the adhesion landscape, cannot capture this set of trends (Fig.~\ref{fig_drugs_adh}). Thus, these results suggest that perturbation of cell polarity, myosin contractility and actin polymerization reduces the strength of the geometry adaptation, indicating that the geometry adaptation mechanism depends on these cellular components.

%------------------------------------------------------------
\section{Discussion}

In this work, we develop a theoretical framework to describe the joint stochastic dynamics of cell nucleus, protrusion, and polarity, and their coupling to the extracellular microenvironment. Experimentally, we find that cells migrating in confinements with a thin constriction exhibit a stereotypical protrusion-nucleus cycling, with characteristic protrusion growth followed by a rapid transition of the nucleus across the bridge. Using a large experimental data set of joint protrusion and nucleus trajectories, we systematically constrain a mechanistic model for confined cell migration. 

In our model, we identify three distinct stages of the protrusion-nucleus cycling (Fig.~\ref{fig_sum}). First, we observe an initial exploration phase, where both nucleus and protrusion are located on the same island (Stage I). At this stage, the polarity is subject to negative feedback, causing the protrusion to frequently change direction and explore its surroundings. Stochastic polarity excitations can trigger the protrusion to enter the constriction. Within the constriction, the protrusion becomes highly confined, causing the polarity dynamics to switch from a negative to a positive feedback loop. This positive feedback reinforces the polarity, driving the protrusion growth into the constriction (Stage II). At the same time, tension builds up due to the coupling to the nucleus, which is held back on the island due to the enhanced adhesion with the substrate. Once the protrusion reaches the other end of the system, the nucleus is pulled across the constriction, relaxing the tension in the elastic coupling, reminiscent of a slingshot (Stage III). The three stages of the transition process arise as a consequence of the interplay of the three key physical mechanisms in the system: the adhesion landscape, the nucleus-protrusion coupling, and the polarity self-reinforcement.

%%%% ADHESION LANDSCAPE %%%%%
To develop this theoretical approach, we separately constrain the dynamics of the nucleus and the protrusion and systematically consider model terms of increasing complexity (Table~\ref{tab:model_table}). We first studied the stochastic dynamics of the cell nucleus. Interestingly, the nucleus dynamics are inconsistent with movement in a simple double-well potential, as might be expected for example from cellular deformation arguments~\cite{Albert2014,Bi2014,Bi2015,Segerer2015,Goychuk2018}. Indeed, active particles confined to double-well potentials can exhibit excitable dynamics similar to those observed in the experimental trajectories of the nucleus alone~\cite{Caprini2019}, making the double well a promising model candidate. However, based on the observed joint dynamics of nucleus and protrusion, we find that this energy barrier model is unable to capture our experiments.

Instead, our model suggests that the movement of the nucleus is determined by the locally available adhesive area, manifesting as an adhesion landscape with a spatially variable friction coefficient. Thus, as the protrusion explores the environment, the back of the cell `sticks' due to the high adhesiveness on the island. This is in line with experimental observations showing that in mesenchymal migration, the movement of the cell rear, where the nucleus is typically located, is limited by the unbinding of mature adhesions~\cite{Gupton2006,Giannone2009}. In the constriction, the cell polarity actively drives the protrusion away from the nucleus, causing mechanical stress to build up in the protrusion-nucleus coupling. This `self-loading' of the coupling eventually causes a contraction stage, where the cell quickly contracts and the nucleus rapidly moves across the constriction. In the model, the tension in the elastic coupling rapidly relaxes during the slip phase, similar to a slingshot. Such `slingshot' dynamics have also been observed in confined 3D migration in fibrous matrices~\cite{Wang2019a}. 

%%%% GEOMETRY ADAPTATION %%%%%
The adhesion-limited nucleus motion is reminiscent of stick-slip processes that have been observed in cell migration on 1D lines~\cite{Monzo2016,Hennig2020,Ron2020a} due to the mechanosensitive binding and unbinding dynamics of adhesions~\cite{Ron2020a,Sens2020}. In contrast, our work suggests that a key determinant for  stick-slip dynamics in confined systems is the interplay of the geometry-sensitive polarity dynamics with the elastic protrusion-nucleus coupling, leading to the self-loading of the coupling. Specifically, the model indicates that the polarity dynamics adapts to the presence of the constricting geometry by activating a self-reinforcing positive feedback loop. This positive feedback leads to a broken-symmetry state, in which there is a non-zero preferred polarity. Symmetry breaking in polarity dynamics has been considered in previous models, including unconfined 2D migration~\cite{Maiuri2015}, chemotaxis~\cite{Prentice-Mott2016}, and protrusion growth in left-right decisions~\cite{Hadjitheodorou2022}. However, in these cases, this state emerged for fixed cell parameters, as a response to chemical concentration, or to resolve competition between protrusions, respectively. In contrast, our work suggests that such states may also arise as a consequence of adaptation to a confinement. 

We were able to rule out an alternative model to the position-dependent feedback where the polarity dynamics do not depend explicitly on external geometry, but on the extension of the cell (Appendix~\ref{sec_strategy}). An interesting aspect of comparing these two models are their conceptually distinct implications. The position-dependent feedback implies a direct coupling to the external environment, where the cell may sense and adapt to the external geometry. In contrast, an extension-dependent feedback is translationally invariant, and only depends on the internal state of the cell. Such a mechanism has recently been suggested to be decisive for protrusion growth in keratocytes~\cite{Raynaud2016a}. In contrast, our findings suggest that the positive feedback loop in the polarity is a response to the geometry of the local microenvironment rather than to the overall extension of the cell. 
%This response to geometry is intimately linked to the local dimensionality of the system: on the islands, the cell polarity can explore a 2D space of directions, while it is effectively contrained to a 1D space in the constriction. 

%%%% MECHANISTIC INTERPRETATION %%%%%
We demonstrate that the geometry adaptation of protrusion and polarity dynamics suggested by our model depends on several underlying biological mechanisms, including the polarity-mediating Rho-associated protein kinase (ROCK), myosin contractility, and actin polymerization. This observation suggests that geometry adaptation may be controlled by the underlying polarization mechanisms of the cell. There are a several ways these mechanisms could contribute to the geometry adaptation of cell polarity dynamics. First, based on the physics of active gels, which describe, for example, the actomyosin cortex in the protrusion, we expect a greater degree of alignment of actin fibers in a narrow constriction~\cite{SoaresESilva2011,Bonelli2016}. Increased alignment of actin is associated with higher myosin-contractility~\cite{Prost2015,Julicher2018} and the emergence of spontaneous cell polarization~\cite{Jiang2005,Prager-Khoutorsky2011,Trichet2012,Gupta2015,Ladoux2016}. A further key determinant of cell polarization are diffusable polarity cues, such as Rho GTPase~\cite{CallanJones2016,Hodge2016,Warner2019}, whose spatiotemporal organization may couple to external geometries, for example through focal adhesions~\cite{Demali2003}, or the cell shape itself~\cite{Maree2012a,Vasilevich2021}. By combining the data-driven mechanistic modelling developed in this work with cytoskeletal perturbations and imaging, the biological and molecular underpinnings of geometry adaptation could be further elucidated in future work.

%%%% MECHANISTIC VS DATA-DRIVEN %%%%%
To make the connection from our model to these molecular processes, microscopic mechanistic models for cell migration could play a key role~\cite{Maiuri2015,CallanJones2016,Recho2019,Ron2020a,Sens2020,Hennig2020,Schreiber2021}. It remains challenging to constrain these models with experimental data. Here, our mesoscopic mechanistic approach could provide a way to bridge this gap. Furthermore, building on these microscopic models could help advance the generalizability of our model by making predictions for more complex confinements, other molecular perturbations and different cell types. Based on experiments in which we varied the dimensions of the constriction as well as the number of adhesive islands to create multi-state micropatterns (Appendix~\ref{sec_geom}), we found that our model already has predictive power beyond the specific confinement geometry used to constrain it. However, determining the adhesion $\gamma(x)$ and polarity feedback landscapes $\alpha(x)$ in more general settings, such as complex geometries, varying protein concentrations or mechanical constraints, may be challenging. These predictions could be complicated by the complex responses of cells to sensory inputs, such as the non-monotonic dependence of cell speed on fibronectin density~\cite{Gupton2006}. Furthermore, we reduced the cellular dynamics to a one-dimensional description, while the effective dimensionality of the dynamics may vary as a function of position in the micropatterns. Generalizing this model to a two-dimensional description could give further insights into how these dynamics are affected by local dimensionality.

Thus, future research is needed to investigate how our model can be generalized and connected with microscopic models to make predictions for new experiments, such as cell migration on patterned lines~\cite{Schreiber2016,Schreiber2021,Caballero2014,Caballero2015}, in 3D-confinements~\cite{Patteson2019,Davidson2020}, or at junctions in a maze~\cite{Renkawitz2019}. Previous work has investigated the effect of asymmetric periodic ratchet-patterns, which led to a rectification of cell migration in one direction~\cite{Mahmud2009,Caballero2014,Caballero2015}. This rectification has been interpreted to be a consequence of the asymmetry in locally available adhesive area~\cite{LoVecchio2020}, consistent with our adhesion landscape model. Our work suggests that the adaptation of cell polarity in response to confinements may also play an important role in such processes. Finally, protrusion and polarity dynamics are critical in migration in 3D extra-cellular matrices~\cite{Friedl2003,Fraley2010,Caswell2018}, as well as in pair-wise interactions of cells~\cite{Abercrombie1953,Carmona-Fontaine2008,Brueckner2021,Zisis2022,LaChance2022,Vercurysse2022}, which in turn control the collective dynamics of cells~\cite{Alert2020}. The geometry adaptation dynamics we have identified here could therefore play an important role in these more complex processes, and provide a new framework for physical models of cell migration in confining systems.

%------------------------------------------------------------
\subsection*{Author Contributions} 
D.B.B., E.H. and C.P.B. conceived the project; A.F., D.B.B., C.P.B. and J.R. designed experiments; A.F. and G.L. performed experiments; A.F., G.L. and N.A. performed tracking; M.S., J.F. and D.B.B. developed the image segmentation; D.B.B. and M.S. analysed data; D.B.B., M.S. and C.P.B. developed the theory; D.B.B. and C.P.B. wrote the manuscript with input from all authors. \\\

\subsection*{Acknowledgements} 
We thank Grzegorz Gradziuk, Steven Riedijk, Janni Harju, and Schnucki for helpful discussions, and Andriy Goychuk for advice on the image segmentation. Funded by the Deutsche Forschungsgemeinschaft (DFG, German Research Foundation) - Project-ID 201269156 - SFB 1032 (Project B12). D.B.B. is a NOMIS fellow supported by the NOMIS foundation and was supported in part by a DFG fellowship within the Graduate School of Quantitative Biosciences Munich (QBM) and by the Joachim Herz Stiftung.

\appendix
\section{Experimental Methods}
\subsection{Sample preparation}
\label{sec:sample_prep}
Fibronectin micropatterns are made by microscale plasma-initiated protein patterning as described previously~\cite{Brueckner2019}. All two-state micropatterns are designed to have adhesive island with square dimensions $((36.7 \pm 0.6)^2  \ \si{\micro\meter}^2)$. For patterns with varying bridge width, we use a standard bridge length $L = 35.3 \pm 0.5 \ \si{\micro\meter}$ and widths $W=3.9 \pm 0.5,6.9 \pm 0.6,12.4 \pm 0.5, 21.7 \pm 0.5,34.8 \pm 0.2 \ \si{\micro\meter}$. For patterns with varying bridge length, we use standard bridge width $W = 6.9 \pm 0.6$ and lengths $L=6.4 \pm 0.3,9.2 \pm 0.3, 23.7 \pm 0.4, 46.2 \pm 0.4, 56.0 \pm 0.3 \ \si{\micro\meter}$. For three-state patterns, a bridge length $L = 24.7 \pm 0.5 \ \si{\micro\meter}$ and width $W=6.9 \pm 0.6$ is used. We refer to the rounded values for $W$ and $L$ throughout the text.

\subsection{Cell culture and microscopy}
MDA-MB-231 cells (DSMZ) are cultured in Minimum Essential Medium (MEM, c.c. pro), containing 10\% FBS (Gibco) and 2mM L-Glutamine (c.c. pro). Cells are grown in a 5\% CO$_2$ atmosphere at 37$^{\circ}$C. For passaging and experiments, cells are washed once with PBS and trypsinised for 3 min. This cell solution is centrifuged at 1000 rcf for 3 min. The cell pellet is re-suspended in MEM and 10,000 cells are added per $\si{\micro}$-dish and left to adhere in the incubator for 4h. The medium is then exchanged to L-15 medium containing L-glutamine (Gibco, supplemented with 10\% FCS) and 25 nM Hoechst 33342 (Invitrogen) for staining cell nuclei. Experiments are performed at 37$^{\circ}$C without CO$_2$. All measurements are performed in time-lapse mode for up to 50 h on an IMIC digital microscope (TILL Photonics) or on a Nikon Eclipse Ti microscope using a 10x objective. The samples are kept in a heated chamber (ibidi GmbH or Okolab) at 37$^{\circ}$C throughout the measurements. Images (brightfield and DAPI) are acquired every 10 mins.

\subsection{Drug treatments} 
To perturb the cells with inhibitor drugs, the corresponding agent is added to the cell culture medium in the usual experimental setup. Y-27632 (Calbiochem/Sigma Aldrich) was added at 2 $\si{\micro}$g/ml, blebbistatin (Cayman Chemical) at 10 $\si{\micro}$M and latrunculin (Merck) at 0.1 $\si{\micro}$M. Blebbistatin and latrunculin concentrations were taken from ref.~\cite{Schreiber2021}. The cells are incubated in the drug containing medium for two to three hours before the start of the measurement to allow the treatment to take effect. The medium is not changed again during the time of measurement.  \\\

\section{Image analysis}
\label{sec_image}

\subsection{Cell segmentation}

The trajectories of the cell nuclei are obtained by applying a band pass filter to the images of the nuclei, binarising, and tracking the binarised images using ImageJ's Analyze Particle plugin~\cite{Schneider2012}. To obain cell shapes, we found that attempts to segment the cell images using traditional methods of image binarization failed and thus turned to more advanced machine learning techniques. Specifically, we utilized convolutional neural networks, which allow for high pixel classification accuracy by accounting for local properties of the image. In particular we use a U-Net architecture, which combines an encoder/decoder structure with skip connections across the latent layers. The encoder/decoder structure allows for efficient recognition of large-scale features in the image, while the skip connections effectively propagate local, low-level information forward in the network. The encoder and decoder branches of our network are three layers deep, with 64 channels in the first layer which are doubled after every max pool layer, similar to previous implementations~\cite{Ronneberger2015}. 

%%%%%%%%%%%%%%%%%%%%%%%%%%%%%%%%%%%%
%FIGURE 
\begin{figure}[h!]
	\includegraphics[width=0.5\textwidth]{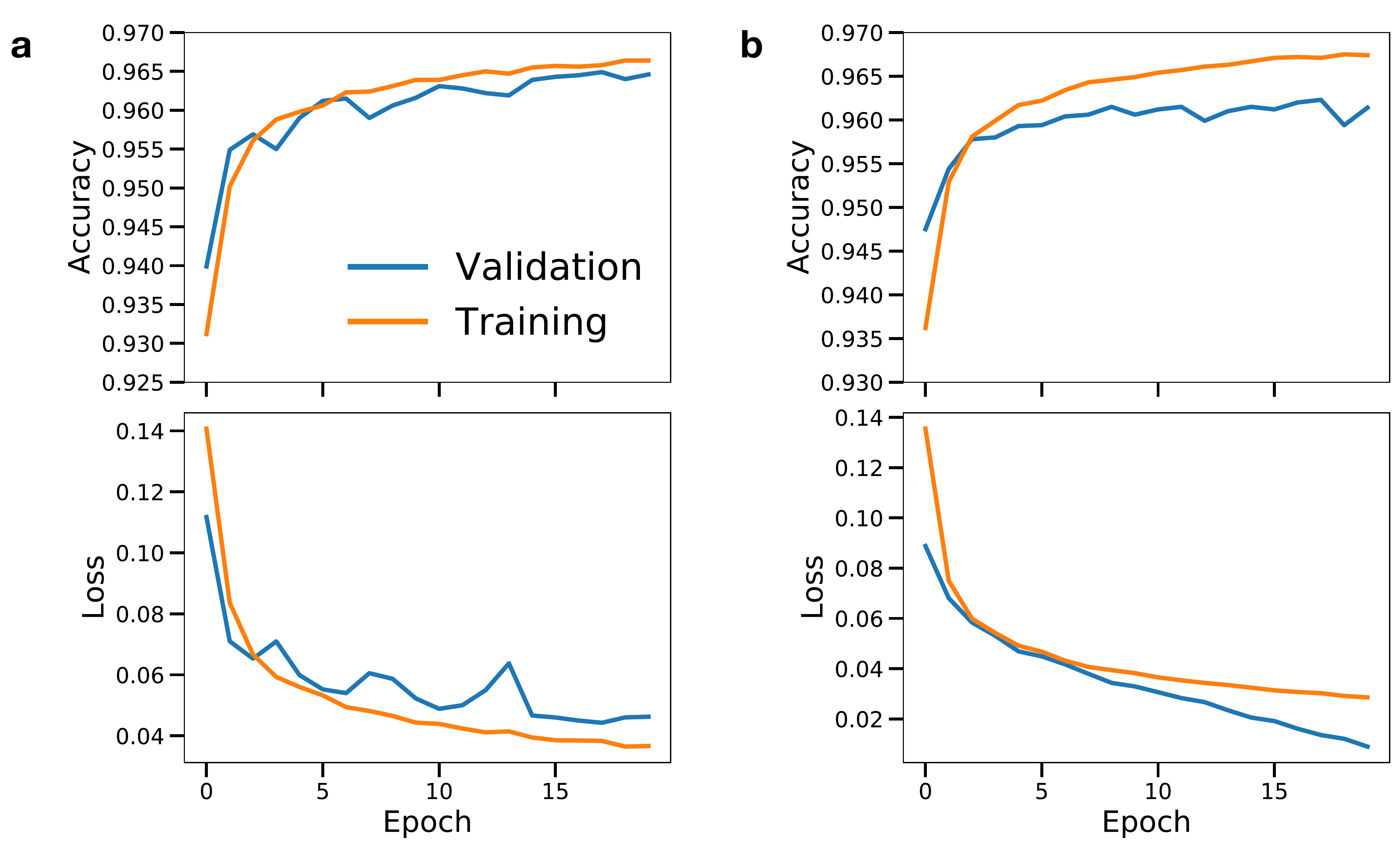}
	\centering
		\caption{
		\textbf{Accuracy and loss curves of the training process for videos with (a) high and (b) low contrast. } 
		The network for high contrast videos was trained with constant $\alpha = 1$, while the network for low contrast videos was trained with varying $\alpha$.
		 }
	\label{losscurves}
\end{figure}
%%%%%%%%%%%%%%%%%%%%%%%%%%%%%%%%%%%%

For training, the network is fed augmented data which has undergone random rotations, shifts, shears, zooms, and reflections. We use $80 \%$ of the original labeled data set of ($N=372$) images for training, and withhold $20 \%$ for validation. Each epoch then consists of 2000 steps of batch size 16, and training is stopped after 20 epochs to prevent overfitting. Gradient updates are performed using the Adam optimizer~\cite{Kingma2015} with a constant learning rate of $10^{-4}$. We used the binary cross-entropy as a loss function to optimize the pixel classification accuracy. For videos with low contrast between the cells and the background, resulting from the use of a different microscope, we adjusted the loss function throughout the training to increase the focus on the cell edges, which improved the segmentation quality, which has been found to have a similar effect in previous work~\cite{Kervadec2019}. Specifically, we use the total loss function
\begin{equation}
\mathcal{L} = \alpha\mathcal{L}_{\text{BCE}} + (1-\alpha)\mathcal{L}_{\text{BCE, edge}}.
\end{equation}
Here, $\mathcal{L}_{\text{BCE}}$ is the binary cross entropy loss for the entire image, and $\mathcal{L}_{\text{BCE, edge}}$ is the binary cross entropy only applied to pixels near the edge of the cell. The factor $\alpha$ is deterministically reduced in each epoch to force the network to specialize and focus on the cell boundary in the later phase of the training, which makes up a comparatively small number of pixels compared to the cell as a whole. The parameter $\alpha$ is initialized to 1 and then gradually reduced by 0.05 with each epoch, which we found improved training compared to a fixed alpha. 

%%%%%%%%%%%%%%%%%%%%%%%%%%%%%%%%%%%%
%FIGURE 
\begin{figure}[h!]
	\includegraphics[width=0.5\textwidth]{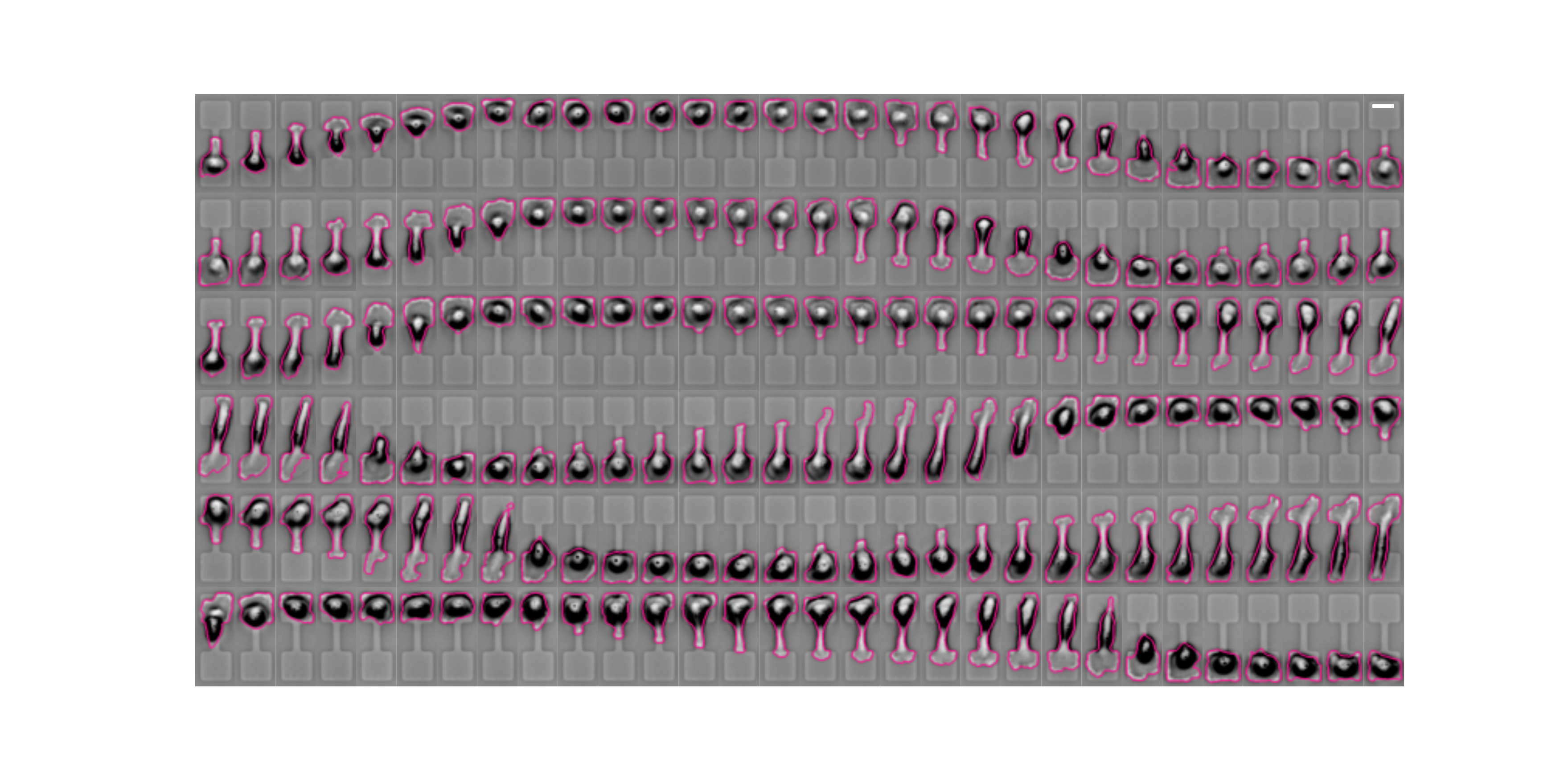}
	\centering
		\caption{
		\textbf{Exemplary brightfield time-series, with segmented cell shapes shown in pink.} 
		Each image is a frame from a video which is sampled every $\Delta t = 10$ min; time flows from the left to the right, and each row is the continuation of the row above it. Bright-field images are inverted for better visibility. Scale bar: $25  \ \si{\micro\meter}$.
		 }
	\label{track_shape}
\end{figure}
%%%%%%%%%%%%%%%%%%%%%%%%%%%%%%%%%%%%

Training according to the above protocol results in a pixel classification accuracy of $96.5 \%$ for videos with high contrast and $96.1 \%$ for videos with low contrast on the validation dataset. We note an apparent slight overfitting, with predictions on the training set achieving a slightly higher accuracy of $96.6\%$ for both high and low contrast videos (Fig.~\ref{losscurves}).

Finally, the predicted segmentations are converted to binary images by applying a threshold. Consequently, pixels with predicted values above 0.12 are mapped to 1, else to 0. This pipeline yields an accurate segmentation of the cell shape for the vast majority of frames (Fig.~\ref{track_shape}). 

\subsection{Protrusion tracking}
\begin{figure}[h!]
	\includegraphics[width=0.5\textwidth]{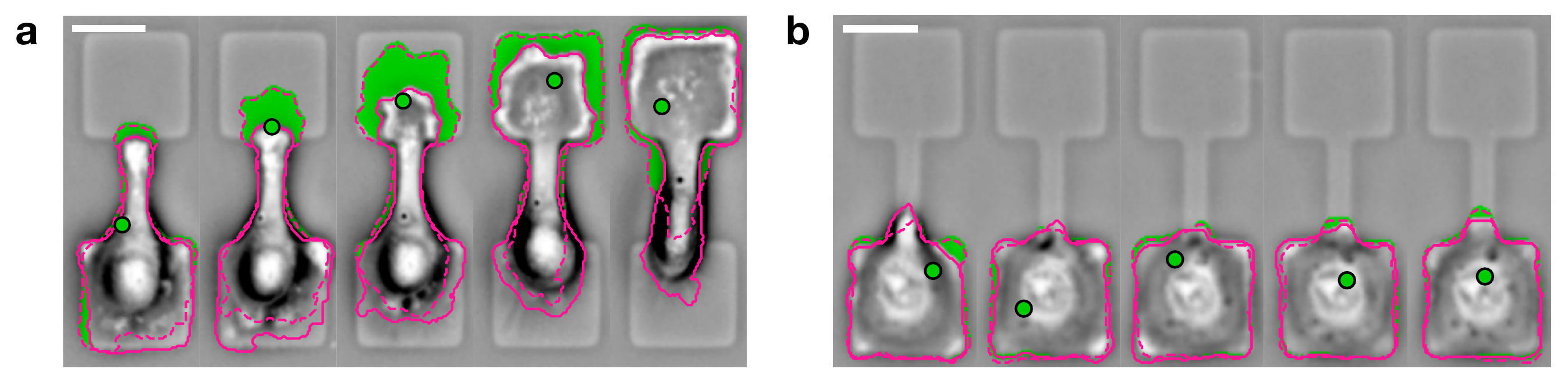}
	\centering
		\caption{
		\textbf{Dynamics of protrusive areas in two frame sequences.} The solid pink line shows the current boundary of the cell area $\mathcal{S}(t)$, and the dashed line is the boundary of $\mathcal{S}(t+\Delta t)$. The protrusive shape (green) is the area which is added between these two frames, $\mathcal{P}(t)$. The geometric center of the protrusive area $\mathbf{x}_\mathrm{p}$ is shown as a green dot. Scale bars: $25  \ \si{\micro\meter}$.
		 }
	\label{track_protrusion}
\end{figure}

To quantify the joint dynamics of nucleus and protrusion motion, we seek a minimal, low-dimensional representation of the cell protrusions. Our image segmentation pipeline gives access to the 2D shape of the cells $\mathcal{S}(t)$ as a function of time. To identify protrusions, we classify the positive contributions to the shape velocities $\mathcal{V}(t) = \mathcal{S}(t+\Delta t)-\mathcal{S}(t)$ as the shape of the protrusion $\mathcal{P}(t)$ (green areas in Fig.~\ref{track_protrusion}).

%%%%%%%%%%%%%%%%%%%%%%%%%%%%%%%%%%%%
%FIGURE 
%%%%%%%%%%%%%%%%%%%%%%%%%%%%%%%%%%%%

%%%%%%%%%%%%%%%%%%%%%%%%%%%%%%%%%%%%
%FIGURE 
\begin{figure}[h!]
	\includegraphics[width=0.5\textwidth]{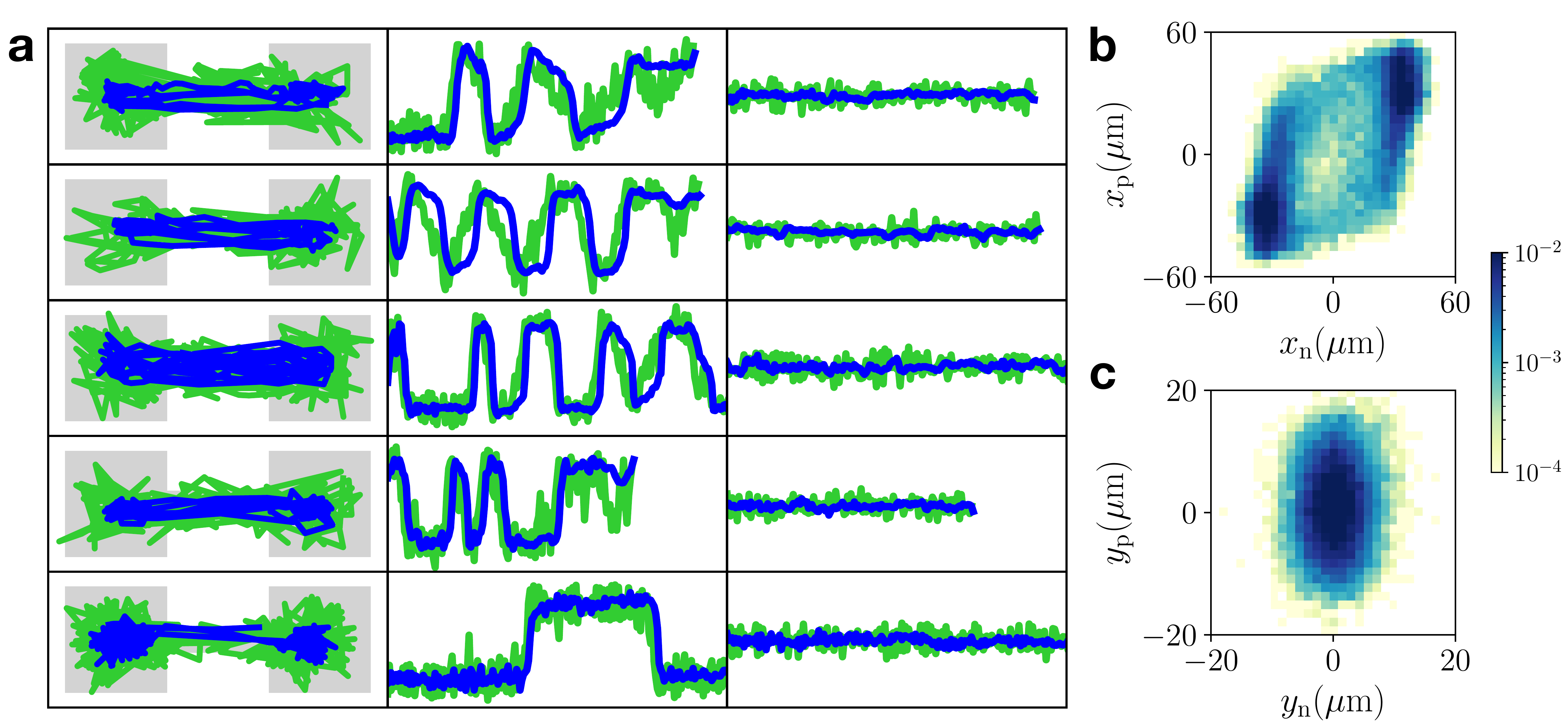}
	\centering
		\caption{
		\textbf{2D motion of nucleus and protrusions.} 
		\textbf{a.} Several examples of 2D trajectories. Left: $xy$-trajectories plotted on top of the micropattern dimension (shown in grey). Axis limits are $-50  \ \si{\micro\meter}<x<50  \ \si{\micro\meter}$ and $-20  \ \si{\micro\meter}<y<20  \ \si{\micro\meter}$; $(x=0,y=0)$ corresponds to the center of the constriction. Middle: $x$-trajectories as a function of time $t$. Axis limits are $-50  \ \si{\micro\meter}<x<50  \ \si{\micro\meter}$ and $0<t<30  \ \si{\hour}$. Right: $y$-trajectories as a function of time $t$. Axis limits are $-50  \ \si{\micro\meter}<y<50  \ \si{\micro\meter}$ and $0<t<30  \ \si{\hour}$, to allow direct comparison with the $x$-trajectories. Blue: nucleus, green: protrusion.
		\textbf{b.} Joint probability distribution $p(x_\mathrm{n},x_\mathrm{p})$ of the $x$-positions, plotted logarithmically. Here shown without the Gaussian interpolation employed in Fig.~\ref{fig1}. 
		\textbf{c.} Joint probability distribution $p(y_\mathrm{n},y_\mathrm{p})$ of the $y$-positions, plotted logarithmically. Note the smaller axis range compared to panel \textbf{b}.
		 }
	\label{xy}
\end{figure}
%%%%%%%%%%%%%%%%%%%%%%%%%%%%%%%%%%%%

As a low-dimensional representation of the protrusive dynamics, we define an effective position of the protrusion $\mathbf{x}_\mathrm{p}$ as the geometric center of the protrusive shape $\mathbf{x}_\mathrm{p}(t) = \int \mathbf{x} \mathcal{P}(t) \d \mathbf{x}$ (green dot in Fig.~\ref{track_protrusion}). The two-state micropattern is designed in such a way that most of the behavior occurs in the $x$-direction along the long axis of the micropattern. Indeed, we find that, similar to the nucleus dynamics, most of the protrusive behaviour is captured by the $x$-component of $\mathbf{x}_\mathrm{p}$ (Fig.~\ref{xy}): the variance in $y$-motion is small (Fig.~\ref{xy}a), and the joint probability distribution $p(y_\mathrm{n},y_\mathrm{p})$ is peaked around $(0,0)$ and exhibits no special structure, unlike the probability distribution for $x$-components $p(x_\mathrm{n},x_\mathrm{p})$ (Fig.~\ref{xy}b,c). In the following, we will therefore take the $x$-component $x_\mathrm{p}$ as a minimal representation of the protrusive dynamics in this system.

\begin{figure}[h!]
	\includegraphics[width=0.5\textwidth]{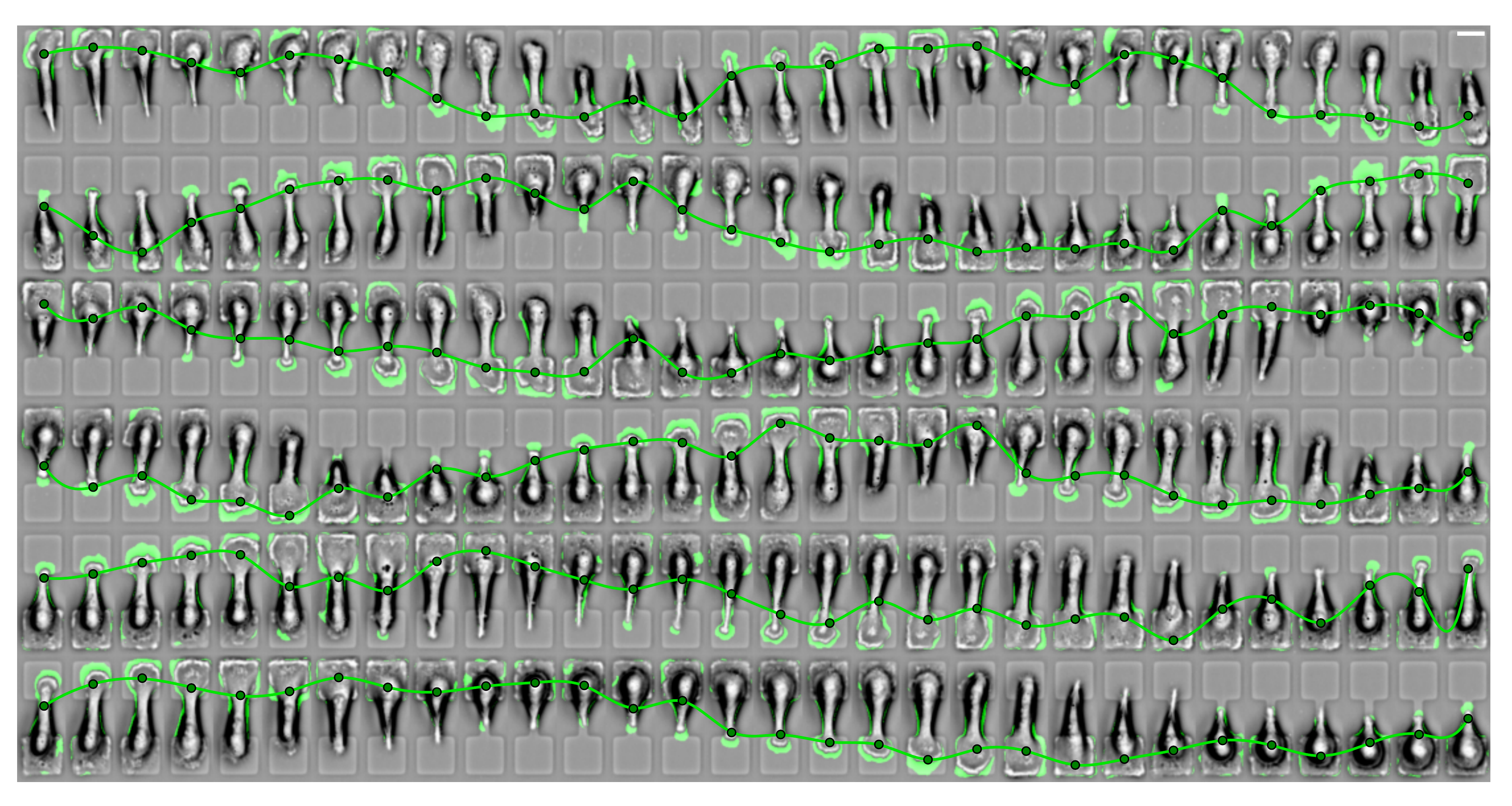}
	\centering
		\caption{
		\textbf{Time series of $x_p$ dynamics overlaid on images of cells with protrusions.} Each image is a frame from a video which is sampled every 10 minutes; time flows from the left to the right, and each row is the continuation of the row above it. Time series curve is an interpolation of the circular points to serve as a guide for the eye.  Scale bar: $25  \ \si{\micro\meter}$.
		 }
	\label{track_protrusion_series}
\end{figure}

We find that this definition captures the characteristic features of the protrusive dynamics during the cell-hopping process: as the protrusion grows into the constriction, the effective protrusion position also moves into the channel (Fig.~\ref{track_protrusion_series} and Fig.~\ref{fig1}d). Thus, $\mathbf{x}_\mathrm{p}$ typically precedes $\mathbf{x}_\mathrm{n}$ in the constriction, as expected from the experimental observations (Supplementary Movies S1-3, Fig.~\ref{fig1}d). Furthermore, we find that  when protrusions form randomly and uniformly around the cell boundary, $x_p$ is located near the cell centroid (Fig.~\ref{track_protrusion}).

%%%%%%%%%%%%%%%%%%%%%%%%%%%%%%%%%%%%
%FIGURE 
%%%%%%%%%%%%%%%%%%%%%%%%%%%%%%%%%%%%

\begin{figure}[h!]
	\includegraphics[width=0.47\textwidth]{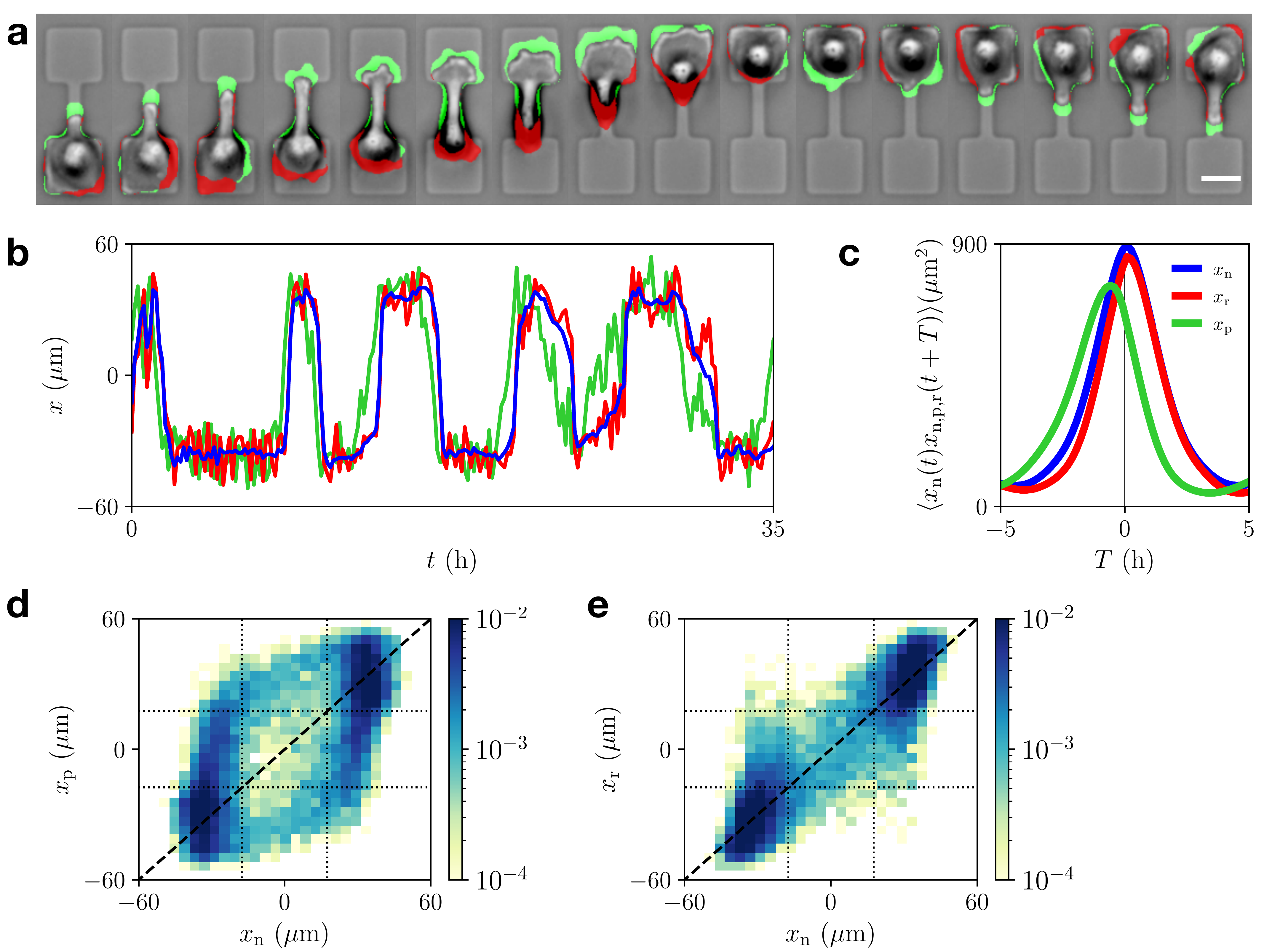}
	\centering
		\caption{
		\textbf{Dynamics of cell retractions.} 
		\textbf{a.} Exemplary brightfield microscopy image series with protrusive shape velocity components $\mathcal{P}(t)$ indicated in green, and retraction components $\mathcal{R}(t)$ in red.
		\textbf{b.} Trajectories of the protrusion $x_\mathrm{p}(t) = \int x \mathcal{P}(t) \d x$ (green), retraction $x_\mathrm{r}(t) = \int x \mathcal{R}(t) \d x$ (red), and the cell nucleus $x_\mathrm{n}(t)$ (blue).
		\textbf{c.} Position cross-correlations between nucleus and protrusion $\langle x_\mathrm{n}(t) x_\mathrm{p}(t+T)\rangle$ (green), between nucleus and retraction $\langle x_\mathrm{n}(t) x_\mathrm{r}(t+T)\rangle$ (red), and nucleus position auto-correlation, $\langle x_\mathrm{n}(t) x_\mathrm{n}(t+T)\rangle$ (blue).
		\textbf{d.} Joint probability distribution $p(x_\mathrm{n},x_\mathrm{p})$ of the $x$-positions of nucleus and protrusion, plotted logarithmically. Here shown without the Gaussian interpolation employed in Fig.~\ref{fig1}. Dashed line indicates the diagonal; dotted lines indicate the boundaries of the adhesive islands.
		\textbf{e.} Joint probability distribution $p(x_\mathrm{n},x_\mathrm{r})$ of the $x$-positions of nucleus and retraction, plotted logarithmically. 
		 }
	\label{retractions}
\end{figure}

In addition to the protrusive dynamics, the cell also performs retractions, corresponding to the negative components of the shape velocities, $\mathcal{R}(t)$ (Fig.~\ref{retractions}a). Using a similar analysis of the retractive dynamics by defining the effective position of the retractions, $x_\mathrm{r}(t) = \int x \mathcal{R}(t) \d x$, we find however that the retractions are well correlated with the position of the nucleus, which typically resides at the rear end of the cell (Fig.~\ref{retractions}b). Specifically, the cross-correlation of nucleus and retraction positions exhibits almost no time-lag, in contrast to the correlation between nucleus and protrusion (Fig.~\ref{retractions}c). Furthermore, the cross-correlation between nucleus and retractions is similar in magnitude and shape to the nucleus position auto-correlation, indicating that the retraction trajectories do not contain significant additional information to the nucleus trajectories. Furthermore, the joint probability distribution of nucleus and retraction positions has maximal probability around the diagonal, with little additional structure, in contrast to the distribution of nucleus and protrusion positions (Fig.~\ref{retractions}d,e). Therefore, to achieve a minimal, low-dimensional description for the coupled dynamics of shape and nucleus motion, we restrict our analysis to the protrusions.

%%%%%%%%%%%%%%%%%%%%%%%%%%%%%%%%%%%%
%FIGURE 
%%%%%%%%%%%%%%%%%%%%%%%%%%%%%%%%%%%%

\begin{table*}[]
\centering
{\begin{tabular}{|c|c||c|c|c|c|c|c|c|c|}
\hline
\multirow{1}{*}{$L$ }& \multirow{1}{*}{$W$ } & \multirow{1}{*}{$L_\mathrm{sys}$} & \multirow{1}{*}{$k_\mathrm{n}$} & \multirow{2}{*}{$\gamma_\mathrm{min}$} & \multirow{1}{*}{$k_\mathrm{p}$} & $\alpha_\mathrm{0}$ & $\alpha_\mathrm{min}$ & $\beta$ & $\sigma$  \\ 
($\mu$m) & ($\mu$m) & ($\mu$m) &  (h$^{-1}$) &  & (h$^{-1}$) & (h$^{-1}$) & (h$^{-1}$) & ($\si{\micro\meter}^{-2} \ \mathrm{h}$) & ($\si{\micro\meter} \ \mathrm{h}^{-3/2}$)  \\ \hline
\multirow{5}{*}{35} & 4 & \multirow{5}{*}{52.5} & \multirow{10}{*}{0.6} & 0.2 & \multirow{10}{*}{1.2} & \multirow{10}{*}{10}  & $-6.5$ & \multirow{10}{*}{$10^{-4}$} & \multirow{10}{*}{100}   \\ \cline{2-2} \cline{5-5} \cline{8-8} 
& 7  &  &  & 0.23 &  &  & $-6$  & &    \\ \cline{2-2} \cline{5-5} \cline{8-8} 
& 12  &  &  & 0.42 &  &  & $-4$  & &    \\ \cline{2-2} \cline{5-5} \cline{8-8} 
& 22  & &  & 0.66 &  &  & 1  & &    \\ \cline{2-2} \cline{5-5} \cline{8-8} 
& 35  & &  & 1 &  &  & 10  & &    \\ \cline{1-3} \cline{5-5} \cline{8-8} 
6 & \multirow{5}{*}{7} & 38 &  & \multirow{5}{*}{0.23} & & & $\multirow{5}{*}{$-6$}$ & &    \\ \cline{1-1} \cline{3-3} 
9 &  & 40  &  &  &  &  & & &    \\ \cline{1-1} \cline{3-3} 
24 &  & 47 &  &  &  &  &  & &    \\ \cline{1-1} \cline{3-3} 
46 &  & 58 &  &  &  &  &  & &    \\ \cline{1-1} \cline{3-3} 
56 &  & 63 &  & &  &  &  & &    \\ \hline
\end{tabular}}
\caption{\label{tab_params} \textbf{Model parameters for varying bridge widths and lengths used throughout the paper.}}
\end{table*}

\section{Model implementation and parameter inference}
\label{sec_implementation}

We implement the spatially variable adhesiveness of the nucleus dynamics (Eq.~\ref{eqn_adhesion}), suggested by the nucleus velocity maps (Fig.~\ref{fig2}), using the dimensionless adhesiveness profile
\begin{align}
\label{eqn_gamma}
\gamma(x_\mathrm{n}) = \frac{1-\gamma_\mathrm{min}}{2} \left(1 - \cos \left(\frac{x_\mathrm{n} \pi }{ L_\mathrm{system}} \right) \right) + \gamma_\mathrm{min}
\end{align}
Thus, $\gamma(x_\mathrm{n})$ varies between $\gamma = \gamma_\mathrm{min}$ at $x_\mathrm{n}=0$ and $\gamma = 1$ on the islands (Table~\ref{tab_params}). The magnitude of the adhesiveness is accounted for the by the parameter $\zeta_\mathrm{n}$. For all length-scale parameters, we use the known dimension of the experimental confinement, i.e. $L_\mathrm{system} = a+L/2 = 52.5 \ \si{\micro\meter}$, where $a$ is the island side length and $L$ the bridge length (see Appendix~\ref{sec:sample_prep}). Thus, Eq.~\eqref{eqn_adhesion} has only two free parameters: $k_\mathrm{n}$ and $\gamma_\mathrm{min}$. We determine these parameters by fitting Eq.~\eqref{eqn_adhesion} to the experimentally observed NVM (Fig.~\ref{fig3}). To constrain the parameters used for all constriction widths throughout, we first fit the thinnest constriction width $W = 4 \ \si{\micro\meter}$, and obtain $k_\mathrm{n} \approx 0.6 \ \mathrm{h}^{-1}, \gamma_\mathrm{min} \approx 0.2$. The fitted value is close to that expected based on purely geometrical arguments: assuming the local friction is proportional to the width of the pattern at that point, we would expect $\gamma_\mathrm{min} \approx W/a \approx 0.1$. The larger actual value could be due to the spatially extended shape of the cell, leading to additional contributions to the adhesive area that are not only determined by the local width of the pattern.

For the protrusion dynamics (Eq.~\ref{eqn_protrusion}), we use soft-wall boundary conditions at the system boundaries, using the potential $V(x_\mathrm{p}) = (x_\mathrm{p}/x_\mathrm{boundary})^{2n}$. Within a reasonable range, the boundary potential parameters do not strongly affect the results; we take $n=4$ and $x_\mathrm{boundary} = 0.4*L_\mathrm{system}$ throughout. Similarly, we find that the choice of $k_\mathrm{p}=k/\zeta_\mathrm{p}$ does not strongly affect the results. Physically, we expect the friction on the nucleus to be larger than on the protrusion, i.e. $\zeta_\mathrm{p}< \zeta_\mathrm{n}$, and thus $k_\mathrm{p}>k_\mathrm{n}$. Fitting the PVM inferred from systems without constrictions, which is dominated by the elastic coupling, we find $k_\mathrm{p}=1.2 \ \mathrm{h}^{-1}$, which we assume to be constant across geometries.

In the geometry adaptation model for the cell polarity (Eq.~\ref{eqn_Kxp}), we use a spatial profile of $\alpha(x_\mathrm{p})$ with a minimal value $\alpha_\mathrm{min}$ at $x_\mathrm{p}=0$ and a maximal value $\alpha_0$ on the adhesive islands:
\begin{align}
\label{eqn_alphaxp}
\alpha(x_\mathrm{p}) = \frac{\alpha_0-\alpha_\mathrm{min}}{2}  \left[ 1+ \cos \left( \frac{x_\mathrm{p} \pi }{ L_\mathrm{system} } \right) \right]
\end{align}
The polarity description has four parameters: $\{ \alpha_0, \alpha_\mathrm{min}, \beta, \sigma \}$. For positions where $\alpha(x_\mathrm{p})<0$, the preferred polarity is $P_0 = \pm \sqrt{|\alpha|/\beta}$. Based on this, we take $\beta=10^{-4} \ \si{\micro\meter}^{-2} \ \mathrm{h}$ throughout, which gives a reasonable order of magnitude of the preferred polarity compared to the typical order of magnitude of the protrusion velocities. On the islands, we assume low protrusion persistence (Supplementary Movie S1), and therefore take $\alpha_0=10 \ \mathrm{h}^{-1}$. This choice yields accurate results for the PVM in systems with no constriction, where we take $\alpha_\mathrm{min}=\alpha_0$, corresponding to a flat profile. Taking smaller values of $\alpha_0$ leads to active driving in the PVM for the system with no constriction, which is not observed experimentally. To constrain $\alpha_\mathrm{min}$, we find that a negative $\alpha_\mathrm{min}$ is required to capture the effective anti-friction in the nucleus dynamics (see Fig.~\ref{fig_MBS} for a parameter sweep). Fitting the anti-friction and dwell time distribution of the thinnest constriction, we obtain $\alpha_\mathrm{min}=-6.5 \ \mathrm{h}^{-1}$. We test the model by increasing $\alpha_\mathrm{min}$ with bridge width, up to $\alpha_\mathrm{min}=\alpha_0$ for the system without constriction (Fig.~\ref{fig_w}, Table~\ref{tab_params}). Finally, the model predictions do not sensitively depend on the choice of the noise amplitude; we take $\sigma=100 \ \si{\micro\meter} \ \mathrm{h}^{-3/2}$ throughout. All parameters are summarized in  Table~\ref{tab_params}.

%%%%%%%%%%%%%%%%%%%%%%%%%%%%%%%%%%%%
%FIGURE 
\begin{figure}[h!]
	\includegraphics[width=0.47\textwidth]{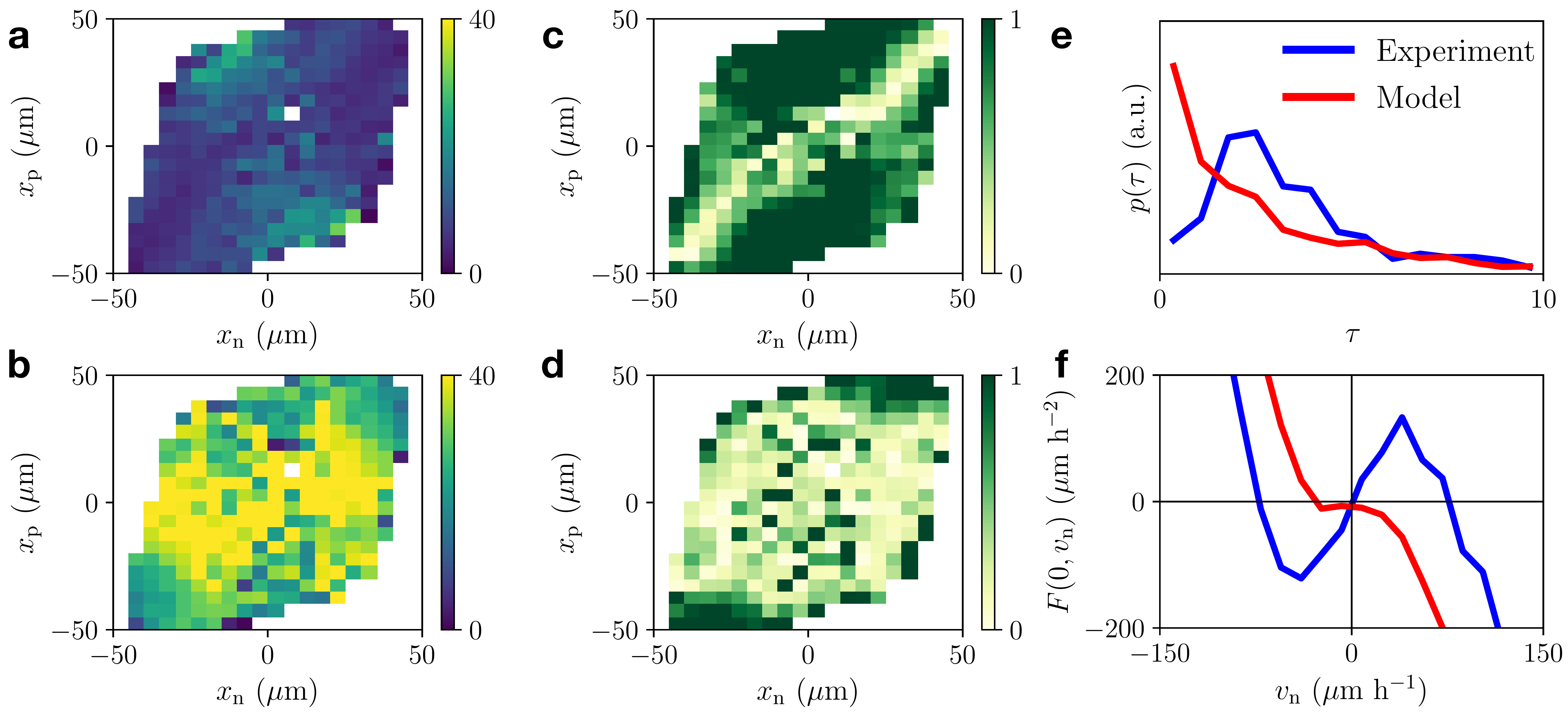}
	\centering
		\caption{
		\textbf{Inferred model terms based on white noise assumption.} 
		\textbf{a.} Inferred multiplicative noise term on the nucleus $\sigma_\mathrm{n}(x_\mathrm{n},x_\mathrm{p}) \approx  (\Delta t \langle [ \dot{x}_\mathrm{n} - f_\mathrm{n}(x_\mathrm{n},x_\mathrm{p}) ]^2 | x_\mathrm{n},x_\mathrm{p} \rangle )^{1/2}$ in units of $\si{\micro\meter} \ \mathrm{h}^{-1/2}$.
		\textbf{b.} Inferred multiplicative noise term on the protrusion.
		\textbf{c.} Relative magnitude of the deterministic and stochastic contributions to the nucleus velocities, for an increment in a time-step $\Delta t$, given by $|f_\mathrm{n}(x_\mathrm{n},x_\mathrm{p})| \sqrt{\Delta t} / \sigma_\mathrm{n}(x_\mathrm{n},x_\mathrm{p})$.
		\textbf{d.} Relative magnitude of the deterministic and stochastic contributions to the protrusion velocities.
		\textbf{e.} Experimental (blue) and predicted (red) dwell time distribution.
		\textbf{f.} Experimental (blue) and predicted (red) effective friction relation $F(x_\mathrm{n} \to 0, v_\mathrm{n})$.
		 }
	\label{inferred_model}
\end{figure}
%%%%%%%%%%%%%%%%%%%%%%%%%%%%%%%%%%%%
%%%%%%%%%%%%%%%%%%%%%%%%%%%%%%%%%%%%%%%%%%%%%%%%%%%%%%%%%%%%%%%%%%%%%%%%
%------------------------------------------------------------
\section{Inferred white noise model does not capture experimental dynamics}
\label{sec_whitenoise}

In this section, we show that a model with a general protrusion term and a white noise polarity dynamics is unable to capture the experimental dynamics. Specifically, we consider a model of the form
\begin{align}
\label{eqn_eom_white_n}
\dot{x}_\mathrm{n} &= f_\mathrm{n}(x_\mathrm{n},x_\mathrm{p}) + \sigma_\mathrm{n}(x_\mathrm{n},x_\mathrm{p}) \xi(t)\\
\label{eqn_eom_white_p}
\dot{x}_\mathrm{p} &= f_\mathrm{p}(x_\mathrm{n},x_\mathrm{p}) + \sigma_\mathrm{p}(x_\mathrm{n},x_\mathrm{p}) \xi(t)
\end{align}
Here, we assume that $\xi(t)$ is a white noise with $\langle \xi(t)\rangle=0$ and $\langle \xi(t)\xi(t') \rangle=\delta(t-t')$. Under this assumption, we can infer the terms $f_\mathrm{n,p}$ and $\sigma_\mathrm{n,p}$ directly from the observed data. Specifically, we use the estimators $f_\mathrm{n}(x_\mathrm{n},x_\mathrm{p}) \approx \langle \dot{x}_\mathrm{n} | x_\mathrm{n},x_\mathrm{p} \rangle$ and $\sigma^2_\mathrm{n}(x_\mathrm{n},x_\mathrm{p}) \approx \Delta t \langle [ \dot{x}_\mathrm{n} - f_\mathrm{n}(x_\mathrm{n},x_\mathrm{p}) ]^2 | x_\mathrm{n},x_\mathrm{p} \rangle$ and similarly for the protrusion terms. These inferred terms provide the best fit estimates for a general model inferred under the white noise assumption. In this case, the inferred functions $f_\mathrm{n}$ and $f_\mathrm{p}$ are given by the NVM and the PVM by definition (shown in Fig.~\ref{fig2}b and Fig.~\ref{fig_Kxp}c). 

The noise on the protrusion significantly exceeds that on the nucleus (Fig.~\ref{inferred_model}a,b). Specifically, the average estimated noise magnitudes are $\hat{\sigma}_\mathrm{n} \approx 8.4 \ \si{\micro\meter} \ \mathrm{h}^{-1/2}$ and $\hat{\sigma}_\mathrm{p} \approx  33 \ \si{\micro\meter} \ \mathrm{h}^{-1/2}$. Accordingly, we find that the nucleus dynamics is dominated by its deterministic component, with deterministic contributions exceeding stochastic fluctuations everywhere in phase space except where nucleus and protrusion are very close together (Fig.~\ref{inferred_model}c). In contrast, the protrusion dynamics are dominated by the stochastic fluctuations (Fig.~\ref{inferred_model}d). In the mechanistic model introduced in the main text, we assume that the source of stochasticity in the system acts on the protrusion, which is further supported by these observations. 

The validity of the model postulated by Eqs.~\eqref{eqn_eom_white_n},~\eqref{eqn_eom_white_p} can be tested by perform simulations using the inferred terms $f_\mathrm{n,p}$ and $\sigma_\mathrm{n,p}$. However, we find that this model fails to predict key experimental observations including the peak in the dwell time distribution and the effective friction (Fig.~\ref{inferred_model}e,f). Taken together, these results indicate that the model postulated in Eqs.~\eqref{eqn_eom_white_n},~\eqref{eqn_eom_white_p} does not provide a good representation of the experimental dynamics, ruling out the white noise protrusion model for all $f_\mathrm{n,p}$.

\section{Ruling out alternative model candidates}
\label{sec_strategy}

To develop a mechanistic model for the joint dynamics of nucleus and protrusion, we systematically constrain the model defined by Eq.~\eqref{eqn_eom_n} and~\eqref{eqn_eom_p}. By employing a data-driven approach based on the conditional averages of the nucleus and protrusion velocities (NVM and PVM respectively), we can independently determine the dynamics of the nucleus and the protrusion. For each of these two components, we systematically increase model complexity step by step until we reach a model that captures the data (Table~\ref{tab:model_table}).

%%%%%%%%%%%%%%%%%%%%%%%%%%%%%%%%%%%%
%FIGURE 

\begin{table}[htb]
\includegraphics[width=0.5\textwidth]{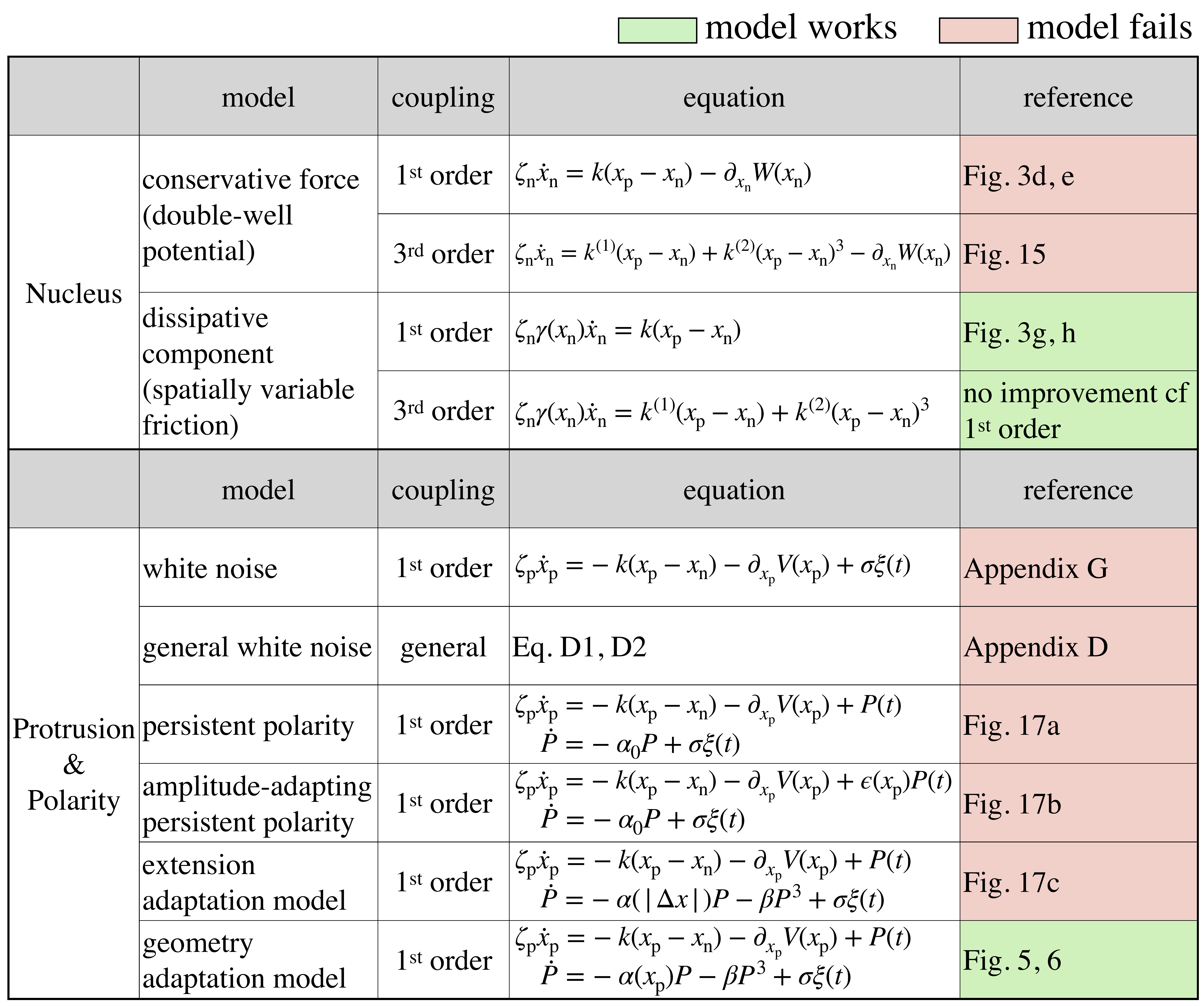}
\caption{
	\textbf{Overview of model candidates.} 
	} 
\label{tab:model_table} 
\end{table}

%%%%%%%%%%%%%%%%%%%%%%%%%%%%%%%%%%%%

%%%%%%%%%%%%%%%%%%%%%%%%%%%%%%%%%%%%%%%%%%%%%%%%%%%%%%%%%%%%%%%%%%%%%%%%
\subsection{Non-linear nucleus-protrusion couplings}
\label{sec_nonlin_coup}
\begin{figure}[h!]
	\includegraphics[width=0.48\textwidth]{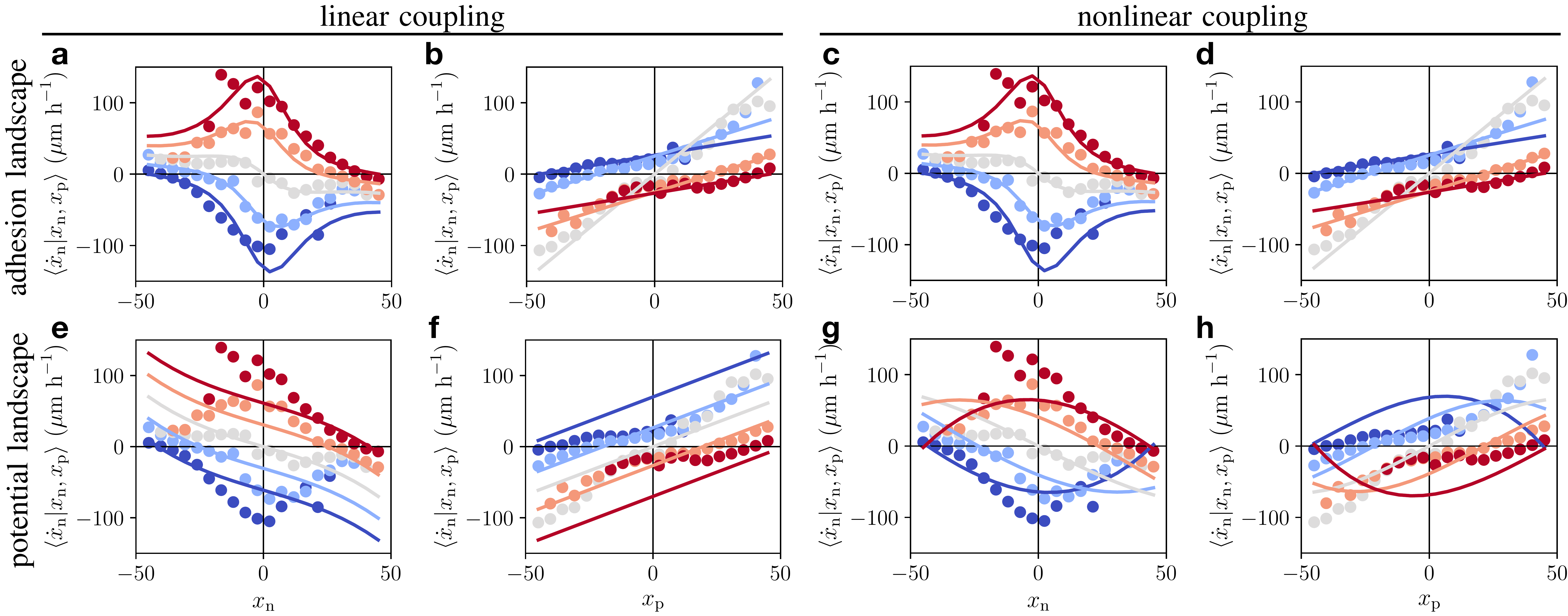}
	\centering
		\caption{
		\textbf{Fitting to the NVM with linear vs non-linear nucleus-protrusion coupling models.} 
		\textbf{a, c, e, g.} NVM as a function of $x_\mathrm{n}$ for different $x_\mathrm{p}$. Dots: Experiment, Line: Fitted linear (panels a, e) or non-linear (panels c, g) coupling model.
		\textbf{b, d, f, h.} NVM as a function of $x_\mathrm{p}$ for different $x_\mathrm{n}$. Dots: Experiment, Line: Fitted linear (panels b, f) or non-linear (panels d, h) coupling model.
		\textit{Top row:} adhesion landscape model. Here, we infer coupling constants $|k^{(2)}_\mathrm{n}|/|k^{(1)}_\mathrm{n}|  \approx 10^{-6}$, indicating that the third order term is negligible compared to the first order term. \textit{Bottom row:} energy potential. 
		 }
	\label{fig_nonlinear}
\end{figure}

We additionally test if a possible nonlinearity in the nucleus-protrusion coupling could provide better predictions with the double-well potential. To this end, we consider the next order coupling term allowed by symmetry in the potential model:
\begin{align}
\label{eqn_nonlin_potential}
\dot{x}_\mathrm{n} = k^{(1)}_\mathrm{n} (x_\mathrm{p}-x_\mathrm{n}) + k^{(2)}_\mathrm{n} (x_\mathrm{p}-x_\mathrm{n})^3 - \partial_{x_\mathrm{n}} W(x_\mathrm{n})
\end{align}
However, this model is unable to capture the NVM features, unlike adhesion model with the first-order coupling (Fig.~\ref{fig_nonlinear} g,h). Furthermore, we show that a nonlinear coupling does not add significant explanatory power to the adhesion model, and is therefore not required to capture our data (Fig.~\ref{fig_nonlinear} a-d).

%%%%%%%%%%%%%%%%%%%%%%%%%%%%%%%%%%%%%
%%FIGURE 
%%%%%%%%%%%%%%%%%%%%%%%%%%%%%%%%%%%%%

%%%%%%%%%%%%%%%%%%%%%%%%%%%%%%%%%%%%%%%%%%%%%%%%%%%%%%%%%%%%%%%%%%%%%%%%
\subsection{Amplitude-adapting persistent polarity}

In the geometry adaptation model, we assume that the feedback on the polarity couples to the external geometry, which therefore makes the time-correlations of the polarity geometry-sensitive. An alternative way to introduce a coupling to the external geometry is a spatially variable overall amplitude $\epsilon(x_\mathrm{p})$. Thus, the time-correlations of this amplitude-adapting polarity $P_\mathrm{AA}$ remain unaffected by the geometry, and only the overall amplitude of the driving force of the protrusion changes. Such a model is described by the equations:
\begin{align}
\label{eqn_POU_xp}
\dot{x}_\mathrm{p} &= -k_\mathrm{p}(x_\mathrm{p}-x_\mathrm{n}) - \partial_{x_\mathrm{p}} V(x_\mathrm{p}) + \epsilon(x_\mathrm{p}) P_\mathrm{AA}(t) \\
\dot{P}_\mathrm{AA} &= - \alpha_0 P_\mathrm{AA} + \sigma \xi(t)
\end{align}
Physically, we expect larger polarities in the constrictions, and therefore employ a generic function $\epsilon(x_\mathrm{p})$ which takes value 1 on the islands and $\epsilon_\mathrm{max}>1$ in the center of the constriction:
\begin{align}
\label{eqn_alphaxp}
\epsilon(x_\mathrm{p}) = \frac{\epsilon_\mathrm{max}-1}{2} \cos \left( \frac{x_\mathrm{p} \pi }{ L_\mathrm{system} } \right) + \frac{\epsilon_\mathrm{max}+1}{2}
\end{align}
Using this implementation, we expect a depletion of probability in the center of the probability distribution $p(x_\mathrm{n},x_\mathrm{p})$ for large $\epsilon_\mathrm{max}$, making it a promising candidate to capture the key features of the protrusion nucleus cycling. To test the model, we screen model predictions across the parameters $\{  \alpha_0, \sigma, \epsilon_\mathrm{max} \}$. Importantly, we find that this model does not capture the experimental observation, including the peaked dwell time-distribution for any combination of parameters (Fig.~\ref{fig_Kdx_xpdx}b).

%%%%%%%%%%%%%%%%%%%%%%%%%%%%%%%%%%%%%
%%FIGURE 
\begin{figure}[h!]
	\includegraphics[width=0.4\textwidth]{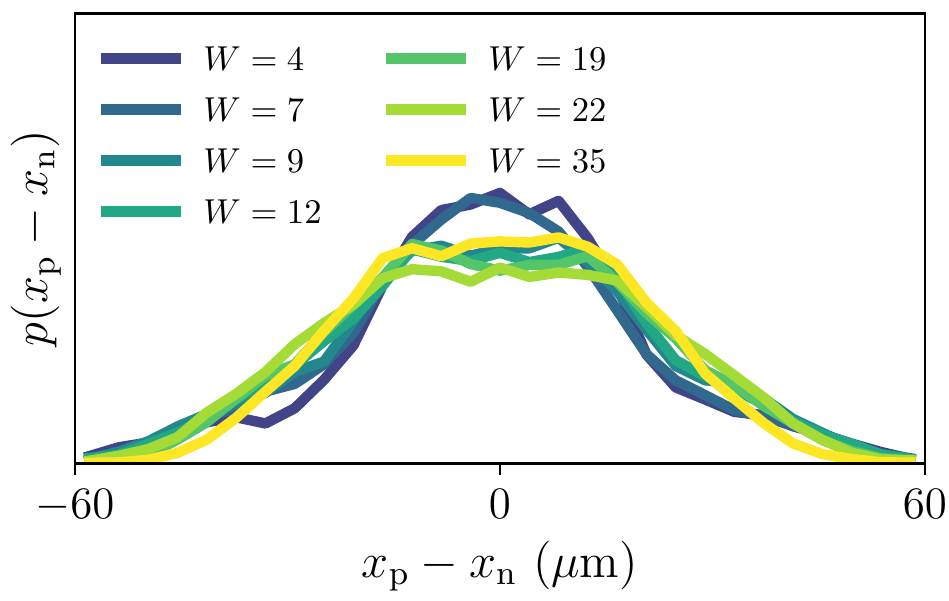}
	\centering
		\caption{
		\textbf{Experimental probability distributions of protrusion-nucleus extension for all bridge widths}
		 }
	\label{fig_pdx}
\end{figure}
%%%%%%%%%%%%%%%%%%%%%%%%%%%%%%%%%%%%%

%%%%%%%%%%%%%%%%%%%%%%%%%%%%%%%%%%%%%%%%%%%%%%%%%%%%%%%%%%%%%%%%%%%%%%%%
\subsection{Extension adaptation polarity}

%%%%%%%%%%%%%%%%%%%%%%%%%%%%%%%%%%%%%
%%FIGURE 
\begin{figure}[b]
	\includegraphics[width=0.47\textwidth]{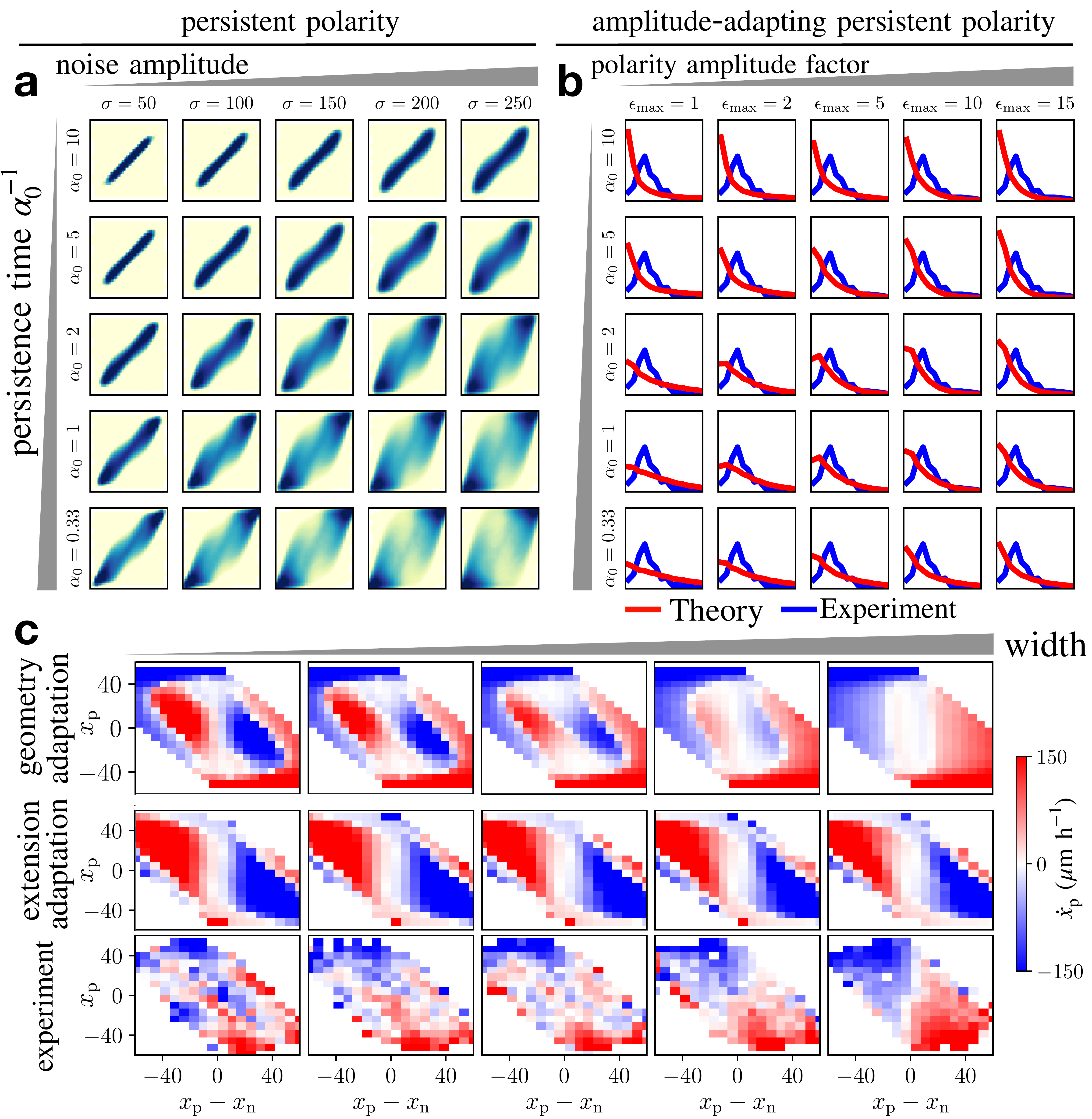}
	\centering
		\caption{
		\textbf{Predictions of ruled-out model candidates.}
		\textbf{a.} Predictions of the persistent polarity with varying noise amplitude $\sigma$ and persistence times $\alpha_0^{-1}$ for the probability distribution $p(x_\mathrm{n},x_\mathrm{p})$. 
		\textbf{b.} Predictions of the amplitude-adapting persistent polarity model with varying polarity amplitude factor $\epsilon_\mathrm{max}$ and persistence times $\alpha_0^{-1}$ for the dwell time distribution. In panels a, b, we use $\gamma_\mathrm{min}=0.23$ and compare to the experimental data for bridge width $W=7 \ \si{\micro\meter}$.
		\textbf{c.} Protrusion velocities as a function of $x_\mathrm{p}$ and $x_\mathrm{p}-x_\mathrm{n}$ for two model candidates. The conditional average $\langle \dot{x}_\mathrm{p} | x_\mathrm{p}-x_\mathrm{n},x_\mathrm{p} \rangle$ is shown for the geometry adaptation model (Eq.~\eqref{eqn_Kxp}, top row), the extension feedback model (Eq.~\eqref{eqn_Kxp_extension}, center row), and the experiment (bottom row), as a function of increasing bridge width left to right.		
		 }
	\label{fig_Kdx_xpdx}
\end{figure}
%%%%%%%%%%%%%%%%%%%%%%%%%%%%%%%%%%%%%

In this section, we show that an alternative model in which the polarity feedback is sensitive to $\Delta x = |x_\mathrm{p}-x_\mathrm{n}|$ instead of the absolute position of the protrusion $x_\mathrm{p}$ is unable to capture our experimental observations. Such a model can be formulated by writing an extension-adapting polarity $P_\mathrm{EA}$ governed by
\begin{equation}
\label{eqn_Kxp_extension}
\dot{P}_\mathrm{EA} =  - \alpha(\Delta x) P_\mathrm{EA} -\beta P_\mathrm{EA}^3 + \sigma \xi(t)
\end{equation}
We expect the polarity to become more persistent for stretched states, with a possible switch to positive feedback at large extensions. As a simple implementation of this dependence, we take $\alpha$ to be a linear function of $\Delta x$:
\begin{align}
\label{eqn_alphadx}
\alpha(\Delta x) = \alpha_0 - \alpha_1 \Delta x
\end{align}
A switch to positive feedback therefore occurs at a critical extension $\Delta x_\mathrm{critical} = \alpha_0/\alpha_1$. Note that since this model does not couple to geometry, $\alpha_0$ and $\alpha_1$ are assumed to be intrinsic cell parameters, which do not adapt to the environment. Therefore, in this model, the bridge width is implemented only through the adhesiveness profile. By contrasting these models, we do not seek to rule out that the cell polarity could couple to the extension of the cell, but investigate whether geometry-sensitive or the extension-dependent polarity dynamics dominate the behaviour in confined cell migration.

%%%%%%%%%%%%%%%%%%%%%%%%%%%%%%%%%%%%%
%%FIGURE 
\begin{figure}[b]
	\includegraphics[width=0.5\textwidth]{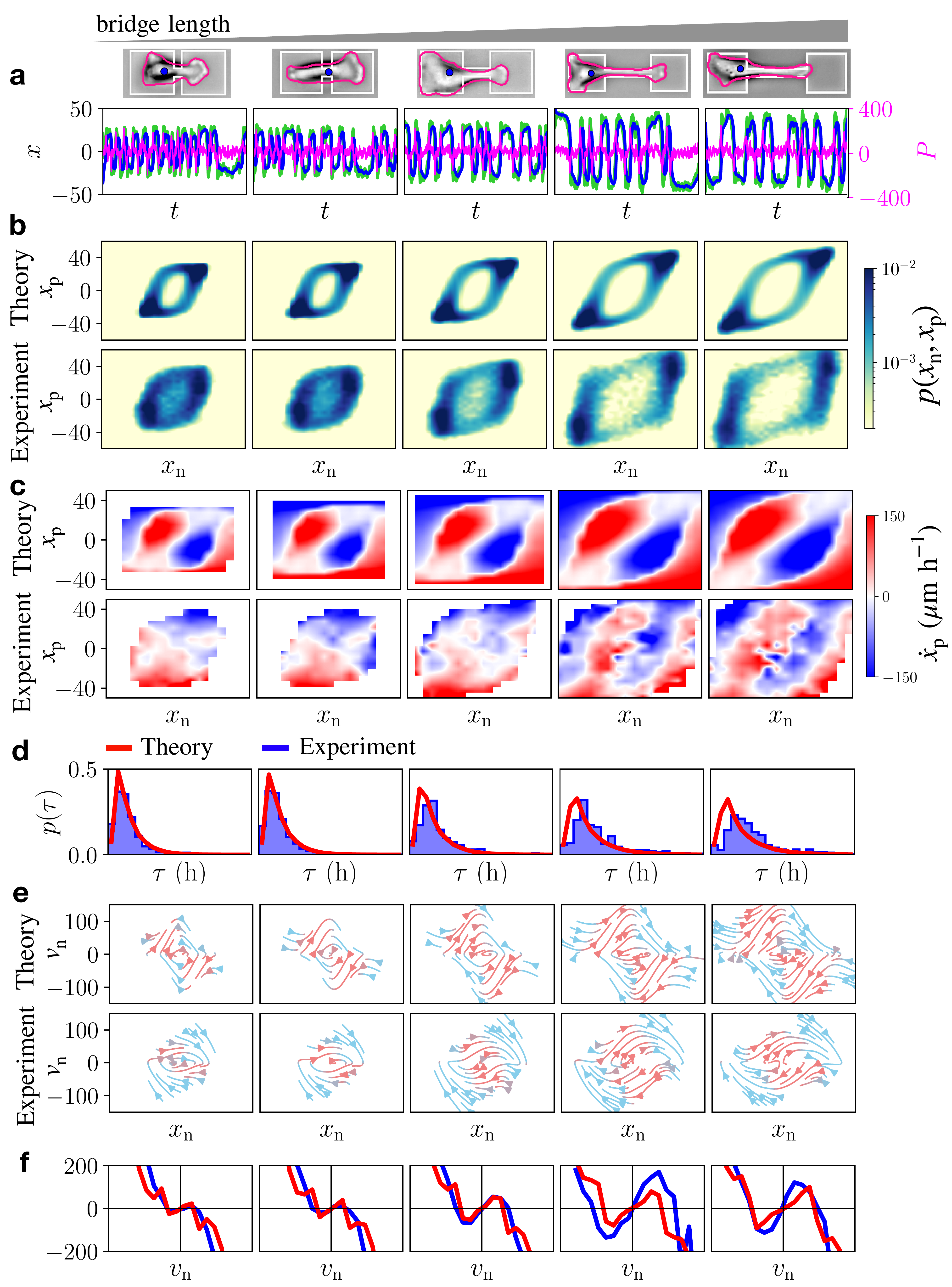}
	\centering
		\caption{
		\textbf{Geometry adaptation model predicts dynamics with varying constriction length.} 
		\textbf{a.} Brightfield microscopy images of MDA-MB-231 cells migrating in geometries with varying constriction length with cell outline in pink and nucleus position in blue, and geometry in white. Bottom: Stochastic trajectories $x_\mathrm{n}(t)$ (blue), $x_\mathrm{p}(t)$ (green), and $P_\mathrm{GA}(t)$ (pink) predicted by the geometry adaptation model.
		\textbf{b.} Joint probability distributions $p(x_\mathrm{n},x_\mathrm{p})$. 
		\textbf{c.} Protrusion velocity maps (PVM) $\langle \dot{x}_\mathrm{p} | x_\mathrm{n},x_\mathrm{p} \rangle$. The top row corresponds to the model prediction, the bottom row to experimental observations.
		\textbf{d.} Predicted (red) and experimental (blue) dwell time distributions $p(\tau)$.
		\textbf{e.} Flow field $(\dot{x}_\mathrm{n},\dot{v}_\mathrm{n})=(v_\mathrm{n},F(x_\mathrm{n},v_\mathrm{n}))$ indicated by arrows~\cite{Brueckner2019}. Arrow color indicates the direction of the local flow: acceleration is orange and deceleration is blue.
		\textbf{f.} Predicted (red) and experimental (blue) effective friction at the bridge center $F(x_\mathrm{n} \to 0,v_\mathrm{n})$.
		In all panels, experimental observations correspond to $L= 6, 9, 24, 46, 56 \ \si{\micro\meter}$ (from left to right).
		}
	\label{fig_L}
\end{figure}
%%%%%%%%%%%%%%%%%%%%%%%%%%%%%%%%%%%%%

In this model, the polarity dynamics has four parameters: $\{ \alpha_0, \alpha_1, \beta, \sigma \}$. We take $\beta=10^{-4} \ \si{\micro\meter}^{-2} \ \mathrm{h}$, $\sigma=100 \ \si{\micro\meter} \ \mathrm{h}^{-3/2}$, and $\alpha_0=10 \ \mathrm{h}^{-1}$, to be consistent with the geometry adaptation model at small extensions (which we have shown to successfully capture the dynamics), and $\alpha_1=1 \ \mathrm{h}^{-1} \si{\micro\meter}^{-1}$, such that the critical extension is $\Delta x_\mathrm{critical} = 10  \ \si{\micro\meter}$; a realistic value given the typical protrusion extensions (Fig.~\ref{fig_pdx}). Interestingly, we find experimentally that the distribution of cell extensions $p(\Delta x)$ does not change significantly with bridge width (Fig.~\ref{fig_pdx}). This suggests that based on the extension feedback model, we expect similar polarity dynamics for all bridge widths, including positive feedback states on all bridge widths, which is in contrast to our the observed nucleus-protrusion dynamics (Fig.~\ref{fig_w}). Indeed, we find that in the parameter regimes where the behaviour in thin constrictions is well captured, the model predictions qualitatively fail to capture the behaviour on wide bridges. This is summarized most clearly in the protrusion velocities as a function of cell extension, which are well captured by the geometry adaptation model, but not by the extension adaptation model (Fig.~\ref{fig_Kdx_xpdx}c). Together, these results rule out the extension adaptation model.

%%%%%%%%%%%%%%%%%%%%%%%%%%%%%%%%%%%%%
%%FIGURE 
\begin{figure*}[t]
	\includegraphics[width=\textwidth]{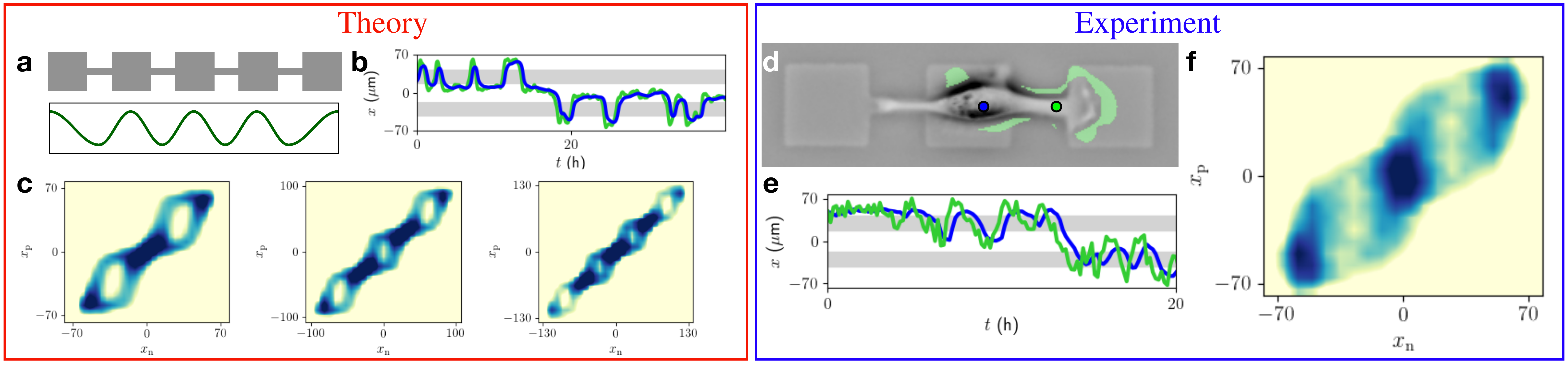}
	\centering
		\caption{
		\textbf{Multi-state micropatterns.}
		\textbf{a.} Sketch of a multi-state micropattern with the corresponding functional form of $\gamma(x_\mathrm{n})$ and $\alpha(x_\mathrm{p})$.
		 \textbf{b.} Model trajectory in a three-state micropattern. Constrictions are marked by grey shades.
		 \textbf{c.} Probability distribution $p(x_\mathrm{n},x_\mathrm{p})$ for 3-, 4- and 5-state micropatterns.
		 \textbf{d.} Microscopy image of an MDA-MB-231 cell migrating in a three-state micropattern, with nucleus and protrusion marked by blue and green dots, respectively. Green area indicates the protrusion area.
		 \textbf{e.} Experimental trajectory in a three-state micropattern. Constrictions are marked by grey shades.
		 \textbf{f.} Probability distribution $p(x_\mathrm{n},x_\mathrm{p})$ for a three-state micropattern.
		 }
	\label{fig_3state}
\end{figure*}
%%%%%%%%%%%%%%%%%%%%%%%%%%%%%%%%%%%%%

%------------------------------------------------------------
\section{Model predictions for varying confinement geometry and drug perturbations}
\label{sec_geom}

To further demonstrate the generality of the model approach, we show how the model can be extended to make predictions for confinement geometries that were not used to constrain the parameters. First, we vary the length $L$ of the constriction (Fig.~\ref{fig_L}a). We find that the model captures the main qualitative changes observed in the experiment (Fig.~\ref{fig_L}b-e), with a `polarity driving' that becomes more strongly pronounced in the longest constrictions in both model and experiment (Fig.~\ref{fig_L}c). Secondly, we test confinements featuring arrays of constrictions, which we term multi-state micropatterns. For such systems, the spatial variations of the adhesion landscape $\gamma(x_\mathrm{n})$ (Eq.~\ref{eqn_gamma}) and the feedback strength $\alpha(x_\mathrm{p})$ (Eq.~\ref{eqn_alphaxp}) are simply extended to periodic functions (Fig.~\ref{fig_3state}a). Using the model parameters constrained for two-state micropatterns, the model then predicts trajectories that oscillate within two-state subsets of the multi-state micropatterns (Fig.~\ref{fig_3state}b). These dynamics lead to probability distribution $p(x_\mathrm{n},x_\mathrm{p})$ that feature concatenations of the two-state probability distributions, with connected ring-like probability distributions (Fig.~\ref{fig_3state}c). To test these predictions, we perform experiments on three-state micropatterns (Supplementary Movie S8, Fig.~\ref{fig_3state}d). We observe trajectories with a similar phenomenology as predicted by the model, and a matching probability distribution $p(x_\mathrm{n},x_\mathrm{p})$.

\begin{figure}[h!]
	\includegraphics[width=0.47\textwidth]{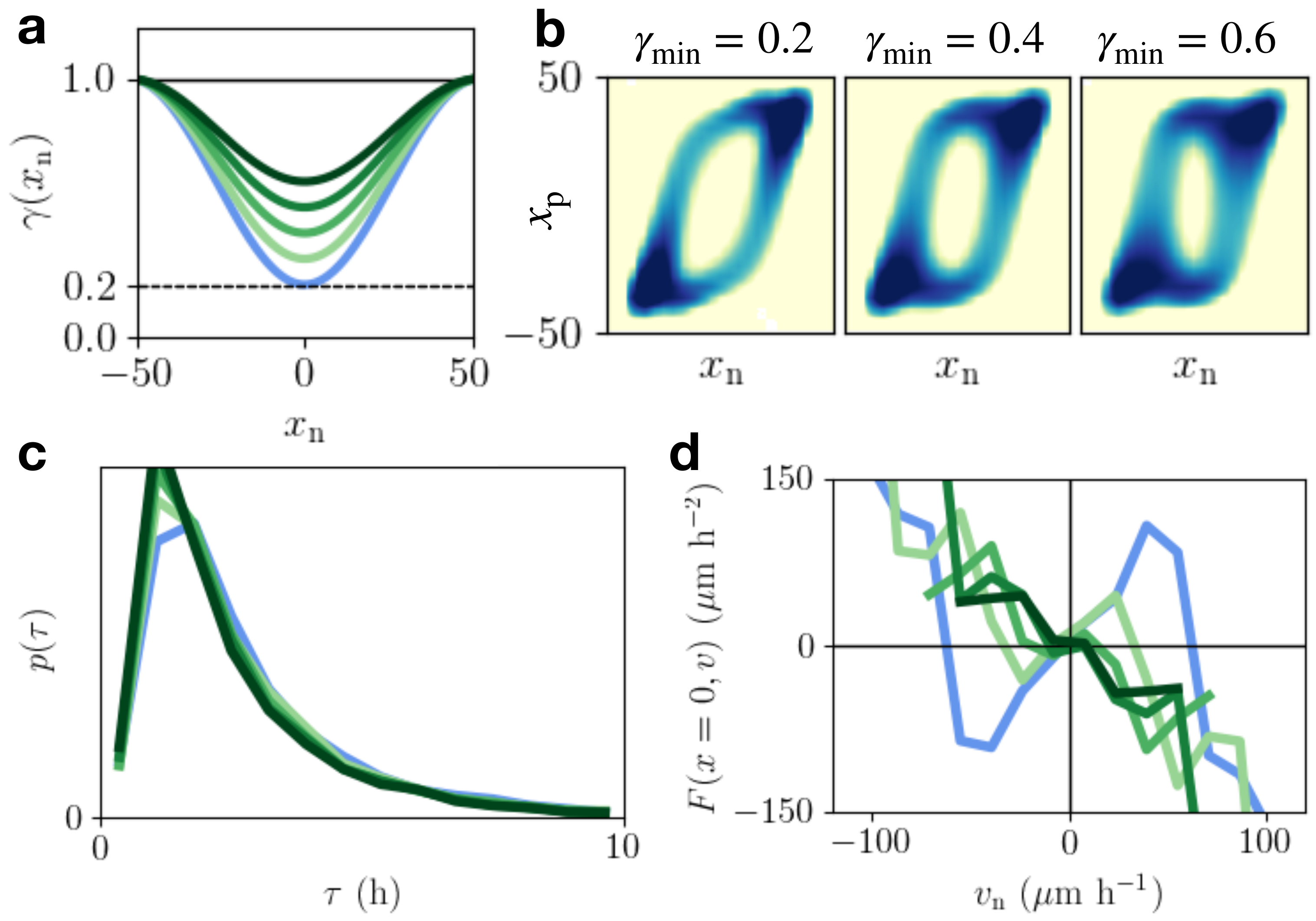}
	\centering
		\caption{
		\textbf{Changes in the adhesion profile upon drug perturbation cannot capture observed changes in nucleus-protrusion dynamics.} 
		\textbf{a.} Adhesion landscape profiles $\gamma(x_\mathrm{n})$ for varying $\gamma_\mathrm{min}=\{0.2,0.3,0.4,0.5,0.6\}$ while keeping the polarity feedback adaptation $\alpha_\mathrm{min}$ constant.
		\textbf{b.} Corresponding probability distributions $p(x_\mathrm{n},x_\mathrm{p})$. The simulations do not predict the closing of the ring-like structure in the probability distribution, as observed experimentally.
		\textbf{c.} Dwell time distributions $p(\tau)$. No significant reduction of the peak in the dwell time distributions is observed, contrary to experiments.
		\textbf{d.} Effective friction at the bridge center $F(x_\mathrm{n} \to 0,v_\mathrm{n})$. The effective friction does exhibit a significant change, but differently to the experiments: Here, the friction switches from a local `negative' friction coefficient in the center of the constriction to a global regular friction $F \sim -v$ (dark green line). In contrast, both the experiments with drug perturbations and the model with reduced geometry adaptation predict a non-linear effective friction with a flat dependence in the constriction center (Fig.~\ref{fig_drugs}i).
				 }
	\label{fig_drugs_adh}
\end{figure}

As a consistency check of our model predictions for perturbations of the geometry adaptation (Fig.~\ref{fig_drugs}), we also make predictions for the consequences of a perturbation of the adhesion landscape. We find that perturbing the adhesion landscape alone cannot capture the effects of the pharmacological perturbations that we tested experimentally (Fig.~\ref{fig_drugs_adh}).

%%%%%%%%%%%%%%%%%%%%%%%%%%%%%%%%%%%%%
%%FIGURE 
%%%%%%%%%%%%%%%%%%%%%%%%%%%%%%%%%%%%%

%%%%%%%%%%%%%%%%%%%%%%%%%%%%%%%%%%%%%
%%FIGURE 
\begin{figure}[b]
	\includegraphics[width=0.5\textwidth]{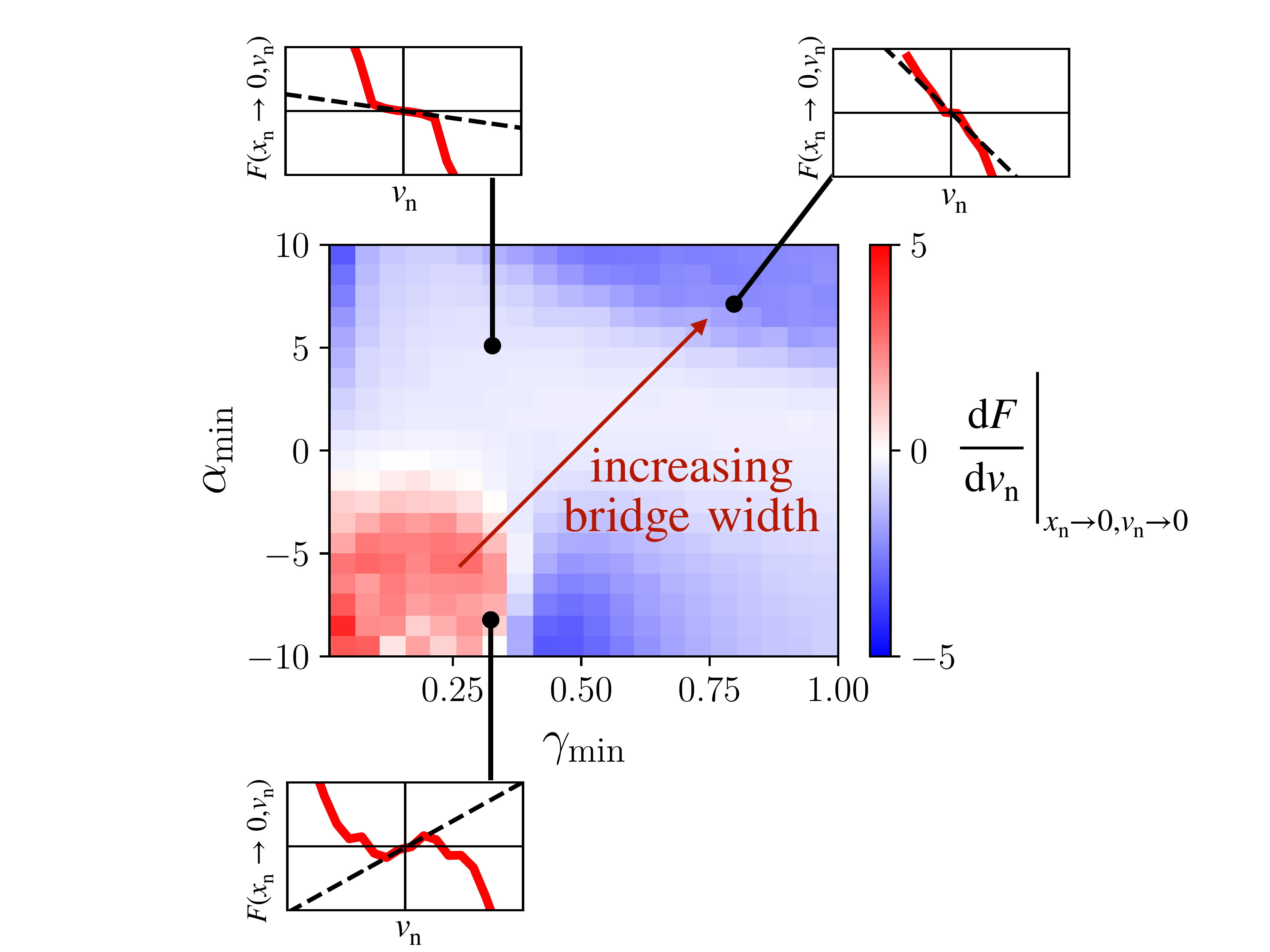}
	\centering
		\caption{
		\textbf{Effective anti-friction for positive feedback polarities.} We vary the parameters $\gamma_\mathrm{min}$ and $\alpha_\mathrm{min}$ and determine the effective friction relationship at the center of the constriction $F(x_\mathrm{n} \to 0,v_\mathrm{n})$. The gradient of the effective friction at $v_\mathrm{n} \to 0$ is indicated by the color. Red arrow corresponds to simultaneously increasing $\gamma_\mathrm{min}$ and $\alpha_\mathrm{min}$ which we do as a model of increasing bridge width. Insets: effective friction relationships at the indicated locations.
		 }
	\label{fig_MBS}
\end{figure}
%%%%%%%%%%%%%%%%%%%%%%%%%%%%%%%%%%%%%

%%%%%%%%%%%%%%%%%%%%%%%%%%%%%%%%%%%%%
%%FIGURE 
\begin{figure*}[t]
	\includegraphics[width=\textwidth]{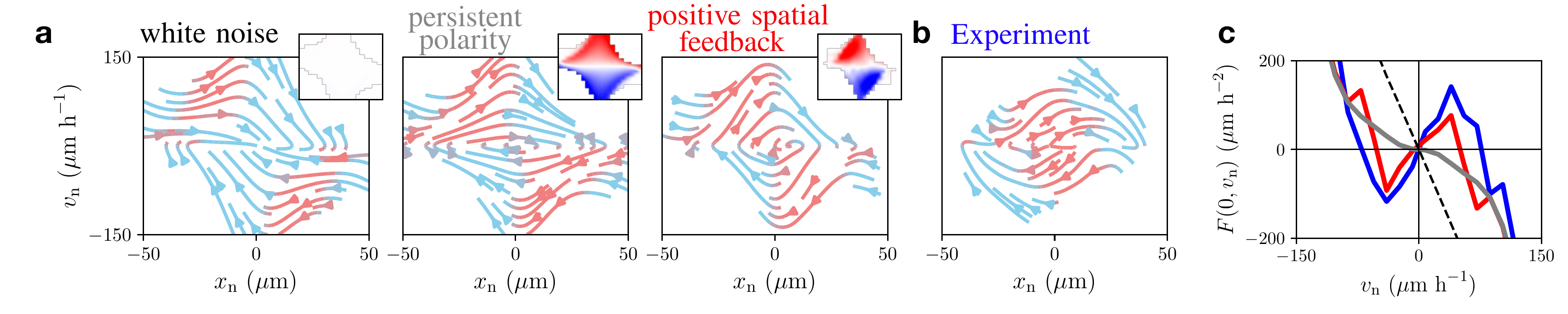}
	\centering
		\caption{
		\textbf{Phase-space portraits of the non-linear dynamics of nucleus trajectories.} 
		\textbf{a.} Flow field $(\dot{x}_\mathrm{n},\dot{v}_\mathrm{n})=(v_\mathrm{n},F(x_\mathrm{n},v_\mathrm{n}))$ indicated by arrows. Arrow color indicates the direction of the local flow: acceleration is orange and deceleration is blue. From left to right, the predictions for the white noise model, the persistent polarity model, and the geometry adaptation model (with $\alpha_\mathrm{min}<0$). \textit{Insets:} conditional average of the polarity as a function of nucleus position and velocity, $\langle P | x_\mathrm{n},v_\mathrm{n} \rangle$, which determines the polarity component of the phase space flow $F_\mathrm{pol}(x_\mathrm{n},v_\mathrm{n})$ (see Eq.~\eqref{eqn_Fxv}).
		\textbf{b.} Flow field inferred from the experiment for bridge width $W=7  \ \si{\micro\meter}$.
		\textbf{c.} Effective friction at the bridge center $F(x_\mathrm{n} \to 0,v_\mathrm{n})$ for all three models and the experiment: white noise model (dashed black line), persistent polarity model (grey), geometry adaptation model (red), and experiment (blue).
		 }
	\label{fig_Fxv}
\end{figure*}
%%%%%%%%%%%%%%%%%%%%%%%%%%%%%%%%%%%%%

%------------------------------------------------------------
\section{Connecting the mechanistic model to emergent stochastic nonlinear dynamics}
\label{sec_xv}

A central challenge for our mechanistic approach is to capture the emergent long time-scale stochastic dynamics of the system. In previous work~\cite{Brueckner2019}, we showed that the stochastic dynamics of the nucleus trajectories $x_\mathrm{n}(t)$ of these cells can be described by an equation of motion for the velocity of the cell nucleus $v_\mathrm{n}$ of the form 
\begin{equation}
\label{eqn_eom_nuc}
\dot{v}_\mathrm{n} = F(x_\mathrm{n},v_\mathrm{n}) + \sigma(x_\mathrm{n},v_\mathrm{n}) \eta(t)
\end{equation}
where $\eta(t)$ is Gaussian white noise, with $\langle \eta(t)\rangle=0$ and $\langle \eta(t)\eta(t') \rangle=\delta(t-t')$. This is an effective description of the dynamics of the nucleus alone, with unobserved degrees of freedom, such as the protrusion and polarity, integrated out. Thus, in contrast to our mechanistic model (Eq.~\ref{eqn_eom_n},~\ref{eqn_eom_p}), the dynamics of the nucleus trajectories alone are described by a \textit{second-order} equation of motion with the velocity $v_\mathrm{n}$ as an additional degree of freedom. Here, we provide a direct mapping between these two descriptions, with the aim to gain insight into how features of the mechanistic dynamics determine the emergent non-linear dynamics of the nucleus motion. Specifically, rewriting Eq.~\ref{eqn_eom_n},~\ref{eqn_eom_p} as
\begin{align}
\label{eom_nuc_rewritten}
\dot{x}_\mathrm{n} &= G_\mathrm{n}(x_\mathrm{n},x_\mathrm{p})  \\
\dot{x}_\mathrm{p} &= G_\mathrm{p}(x_\mathrm{n},x_\mathrm{p}) + P(t)
\end{align}
we can recast these equations into a single differential equation for $v_\mathrm{n}$ by differentiation of Eq.~\eqref{eom_nuc_rewritten}. Then, using the definition $F(x_\mathrm{n},v_\mathrm{n}) = \langle \dot{v}_\mathrm{n} | x_\mathrm{n},v_\mathrm{n} \rangle$, we find
\begin{equation}
\label{eqn_Fxv}
F(x_\mathrm{n},v_\mathrm{n}) = \underbrace{ G_\mathrm{p} \frac{\partial G_\mathrm{n}}{\partial x_\mathrm{p}} + v_\mathrm{n} \frac{\partial G_\mathrm{n}}{\partial x_\mathrm{n} } }_{F_\mathrm{cc}(x_\mathrm{n},v_\mathrm{n})} + \underbrace{ \frac{\partial G_\mathrm{n}}{\partial x_\mathrm{p}} \langle P | x_\mathrm{n},v_\mathrm{n} \rangle }_{F_\mathrm{pol}(x_\mathrm{n},v_\mathrm{n})}
\end{equation}
Here, the right hand-side can be turned into an equation of $x_\mathrm{n},v_\mathrm{n}$ only by replacing $x_\mathrm{p}$ with the inverse of Eq.~\eqref{eom_nuc_rewritten}. Thus, we expect the deterministic dynamics of the nucleus to be determined by two components. A component $F_\mathrm{cc}(x_\mathrm{n},v_\mathrm{n})$ determined by the confinement and coupling dynamics, and a component $F_\mathrm{pol}(x_\mathrm{n},v_\mathrm{n})$ determined by the polarity dynamics.

For white noise polarities, the second term vanishes, as $\langle P | x_\mathrm{n},v_\mathrm{n} \rangle = 0$ (Insets Fig.~\ref{fig_Fxv}a), and thus the phase space flow is due to the combined effects of nucleus-protrusion coupling and the space-dependent adhesiveness acting on the nucleus. Interestingly, for the white noise model, we find a small region of deterministic amplification where the nucleus enters the constriction - however, the amplification only sets in at high speeds, while there is no amplification for low speeds (Fig.~\ref{fig_Fxv}a). This amplification in the flow is due to the differential adhesiveness, as it vanishes for a flat adhesiveness profile. In contrast, in the experiments, we found that the excitable amplification regime sets in already at low speeds. Furthermore, the effective friction acting on the nucleus in the white noise model is a simple linear friction, $F(x_\mathrm{n} \to 0,v_\mathrm{n}) \propto -v_\mathrm{n}$, in contrast to the non-linear anti-friction observed experimentally (Fig.~\ref{fig_Fxv}b). 

Persistence of the polarity leads to a significant contribution to the deterministic dynamics, with $\langle P | x_\mathrm{n},v_\mathrm{n} \rangle$, leading to amplification even at small velocities. However, while the persistent polarity model predicts a non-linear effective friction relation $F(x_\mathrm{n} \to 0,v_\mathrm{n})$, it does not predict a sign-change, corresponding to anti-friction, in any parameter regime we investigated (Fig.~\ref{fig_Fxv}c). In contrast, we find that the geometry adaptation model captures the effective anti-friction at the center of the constriction. In the parameter regime relevant to the experiments, we find that the effective anti-friction emerges for $\alpha_\mathrm{min}<0$ and $\gamma_\mathrm{min} \lesssim 0.3$ (Fig.~\ref{fig_MBS}). To model the effects of increasing constriction width, we simultaneously increase $\gamma_\mathrm{min}$ and $\alpha_\mathrm{min}$ (red arrow Fig.~\ref{fig_MBS}). We observe that this leads to the disappearance of the effective anti-friction, first giving rise to a flat non-linear friction, and finally an almost linear regular friction (Insets Fig.~\ref{fig_MBS}). We observe very similar changes in the effective friction in the experiment (Fig.~\ref{fig_w}).

In summary, the effective non-linear dynamics of the nucleus trajectories put strong constraints on the mechanistic model, and in contrast to the white noise and persistent polarity models, the geometry adaptation model is able to capture the experimentally observed dynamics. The mechanistic approach furthermore gives insight into the origin of the non-linear dynamics: $F(x_\mathrm{n},v_\mathrm{n})$ is composed of a confinement-coupling and a polarity component. The effective anti-friction exhibited by the inferred dynamics is reproduced for parameters corresponding to positive polarity feedback, indicating that such a feedback mechanism may be required to explain the emergence of effective anti-friction in the underdamped nuclear dynamics.

\bibliography{/Users/D.Brueckner/Documents/Mendeley_Desktop/library}

\end{document}